\documentclass[aps,pra,superscriptaddress,floatfix,twocolumn,longbibliography]{revtex4-2}
\usepackage[T1]{fontenc}
\usepackage[utf8]{inputenc}
\usepackage{xcolor}
\usepackage{amsmath}
\usepackage{amssymb}
\providecommand{\coloneqq}{\mathrel{\mathop:}=}
\usepackage{graphicx}
\usepackage{esint}
\PassOptionsToPackage{normalem}{ulem}
\usepackage{ulem}
\usepackage[bookmarks=false,
 breaklinks=false,pdfborder={0 0 1},colorlinks=true]
 {hyperref}
\hypersetup{
 citecolor=blue,urlcolor=magenta}

\makeatletter
\newcommand{\VILNIUS}{Institute of Theoretical Physics and Astronomy, Faculty of Physics, Vilnius University, Saul\.etekio 3, LT-10257, Vilnius, Lithuania}

\makeatother

\begin{document}
\title{Aharonov-Casher Chern bands for ultracold dark state atoms}

\author{Domantas Burba}
\email{domantas.burba@ff.vu.lt}
\affiliation{\VILNIUS}
\author{Dominykas Borodinas}
\affiliation{\VILNIUS}
\author{Gediminas Juzeli\={u}nas}
\email{gediminas.juzeliunas@tfai.vu.lt}
\affiliation{\VILNIUS}

\date{\today}% It is always \today, today,

\begin{abstract}

We consider the Aharonov-Casher (AC) condition for ultracold atoms adiabatically following the dark-state in a $\Lambda$-type atom-light coupling scheme. The AC condition establishes a relation between the geometric scalar potential and the synthetic magnetic field, resulting in a fully degenerate lowest Landau-level-like band in the adiabatic limit even if the magnetic field is inhomogeneous but has a proper sign. 
We derive a general requirement for the atom-light coupling under which the AC condition applies. The requirement holds if the Rabi frequencies of the $\Lambda$ scheme are superpositions of plane waves with the appropriate amplitudes and phases. In particular, the Rabi frequencies made of $N=3,\,4,\,6$ fine tuned plane waves yield a smooth background magnetic field of definite sign, as well as an array of non-measurable Aharonov-Bohm flux singularities. Departing from the fine tuning, the latter singularities turn into narrow subwavelength patches of the magnetic field with an opposite sign, which broaden the lowest energy band. The lowest band is also broadened for the fine tuned situation due to deviation from the adiabatic approach because of the finite atom-light coupling strength. It is shown that a proper combination of the two imperfections (departure from fine tuning and finite atom-light coupling strength) can lead to a significant reduction of the width of the lowest band, which can be made almost completely flat and furthermore becomes characterized by a nearly perfect geometry needed for simulating the fractional Hall states.

\end{abstract}

\maketitle

\section{Introduction}

The quantum-mechanical motion of a charged particle in a homogeneous
magnetic field is characterized by degenerate Landau levels that play
an essential role in the integer and fractional Quantum Hall effects
\cite{Ezawa11QHE-book}. The degeneracy of Landau levels generally
disappears for an inhomogeneous magnetic field when the strength of the
magnetic field pointing along the $z$ axis depends on the transverse
coordinates $x$ and $y$. In 1979 Aharonov and Casher proved that
a spin-1/2 charged particle moving in the $xy$ plane with a spin
pointing along such an inhomogeneous magnetic field has $r-1$ degenerate
ground states with zero-energy, where $r$ is an integer closest to
the total flux in units of the flux quantum \cite{Aharonov-Casher79PRA}.
Thus, the lowest Landau level remains degenerate for the non-homogeneous
magnetic field. 

Recently, it was noticed that a situation similar to that considered
by Aharonov and Casher can arise for the adiabatic motion of the valence
band holes in condensed matter bilayer systems \cite{Shi24PRB} or
for the adiabatic motion of ultracold atoms in  laser fields \cite{Sommer-Cooper2509.01481}.
This relies on the fact that the adiabatic motion of a quantum particle
is affected by the geometric scalar and vector potentials, as well
as the corresponding effective magnetic field \cite{Mead1992,Dalibard11RMP,Goldman2014}.
By properly choosing the parameters of the system, the effective magnetic
field and the scalar potential can obey the Aharonov-Casher (AC) condition
\cite{Shi24PRB,Sommer-Cooper2509.01481} leading to the degeneracy
of the lowest Landau level, which is important for simulating the
fractional Quantum Hall effect. In particular, it was shown that for
ultracold atoms the AC conditions can be achieved in the $\Lambda$
scheme of atom-light coupling involving three plane waves  \cite{Sommer-Cooper2509.01481},
\footnote{In ref.~\cite{Sommer-Cooper2509.01481} it was noted that the AC condition is also satisfied for Rabi frequencies involving four symmetrically distributed fine tuned plane waves used in the revised version of ref.~\cite{Nascimbene25PRL}.}.
In related recent developments, the adiabatic motion of the dark-state
atoms of the $\Lambda$ scheme was considered to create a non-staggered
magnetic field and the associated narrow Landau levels \cite{Nascimbene25PRL,Burba25PRR}.

Synthetic gauge fields that affect the adiabatic motion of dark
state atoms have received considerable interest over the last couple
of decades  \cite{Dum96PRL,VisserNienhuis1998,Juz05JPB,Juz05PRA,Ruseckas05PRL,Dalibard11RMP,Goldman2014}.
This can serve
as a means to create the synthetic magnetic field acting on ultracold atoms \cite{Juz04PRL,Juz05JPB,Juz05PRA,Gvozdiovas23PRA,Nascimbene25PRL,Burba25PRR,Sommer-Cooper2509.01481} or to form narrow subwavelength barriers for ultracold atoms in one dimension
\cite{Zoller2016,Jendrzejewski2016,Wang2018,Zubairy2020,Kubala2021,Gvozdiovas2021}
and two dimensions \cite{Gvozdiovas23PRA}. In the latter two dimensional
(2D) case, in addition to narrow subwavelength potential barriers,
the synthetic magnetic field appears \cite{Gvozdiovas23PRA}, which
can be made non-staggered by reducing to zero the width of selected 2D barriers \cite{Burba25PRR}. Note that in the initial developments of the light-induced synthetic magnetic field using the $\Lambda$ configuration of the atom-light coupling \cite{Juz04PRL,Juz05JPB,Juz05PRA}, the  atoms were considered to interact with vortex light beams, such as Laguerre-Gauss or Bessel beams. In that case, the number of Dirac flux quanta constituting the magnetic flux can not exceed the vorticity of the light beams involved. Much larger magnetic fluxes can be created if one of the light beams acting on atoms in the $\Lambda$ coupling scheme represents an array of vortices and antivortices, which can be created, for example, using standing waves propagating in the orthogonal directions with the $\pi/2$ phase difference \cite{Gvozdiovas23PRA,Nascimbene25PRL,Burba25PRR}. In that case, the magnetic field can be made non-staggered by properly choosing the second light beam  \cite{Nascimbene25PRL,Burba25PRR}. 

Here we provide a general framework for achieving the AC conditions
for the adiabatic motion of the dark state atoms.   This can be
done if the laser radiation inducing atomic transitions in the $\Lambda$
atom-light coupling scheme is made of  properly chosen   plane
waves.  In particular, by using three, four, or six fine-tuned plane
waves, the magnetic field not only obeys the AC criterion, but can
also be smooth and non-staggered,  making the lowest Landau level
completely flat in the adiabatic limit, i.e. for a sufficiently large strength of the atom-light coupling.  We show that an arbitrarily small deviation of the system parameters from this fine-tuned
situation leads to the formation of a periodic array of narrow subwavelength
peaks of a strong magnetic field with a sign opposite to that of the
smooth background field. The peaks are formed at the zero points of the atom-light coupling corresponding to fine tuning of laser fields. Away from
these subwavelength patches of the strong magnetic field, a smooth
magnetic field of the opposite sign remains, compensating for the
former peaks, so the total magnetic flux over an elementary cell is
zero.  As the widths of the subwavelength peaks of the magnetic fluxes
are sufficiently small, their influence is minimal. In that case the
problem is close to the motion of a particle in a non-staggered magnetic
field obeying the AC condition, yet the lowest energy level acquires some finite width. The lowest band is also broadened in the case of fine tuning of laser fields due to deviation from the adiabatic approach because of the finite atom-light coupling strength. It is shown that a proper  combination of the two imperfections (departure from fine tuning  and finite atom-light coupling strength) can lead to the significant narrowing of the lowest energy band, which becomes almost completely flat and furthermore acquires a nearly perfect geometry needed for simulating the fractional Hall states if the atom-atom interaction is added. 

The paper is organized as follows. In the next Section~\ref{sec:Hamiltonian}
we introduce a Hamiltonian for an atom interacting with the laser
radiation in the $\Lambda$ scheme of the atom-light coupling and
define the dark states which are immune to the atom-light coupling.
Section~\ref{sec:Adiabatic-motion-in} presents a general description
of the adiabatic motion of the dark state atoms affected by the geometric
scalar and vector potential, as well as the corresponding magnetic
field. In Section~\ref{sec:Perfect-Chern-insulators} we provide
a general analysis of the two-dimensional (2D) motion of a particle
in the magnetic field, subsequently exploring the applicability of
the AC condition for the adiabatic motion of the dark state atoms
of the $\Lambda$ scheme. Specifically, it is shown that the AC condition
holds if one of the Rabi frequencies of the $\Lambda$ scheme is
a superposition of plane waves characterized by wave-vectors with equal
lengths, whereas another Rabi frequency is a  superposition of
plane waves with opposite wave-vectors and properly chosen amplitudes
and phases.  In Section~\ref{sec:Symmetrically distributed plane waves}
the general formalism is applied to the gauge fields produced by three, four and six symmetrically distributed plane waves ($N=3,\,4,\,6$) providing the periodic atom-light coupling. The quasi-crystaline structure
produced by five plane waves is also considered. Section~\ref{sec:Eigenstates-and-spectrum}
provides the calculation of the spectra for the periodic atom-light coupling for $N=3$ and $N=4$, including the chiral edge states for the cylindrical geometry, the bulk energy spectrum and the band topology. In particular, it is shown that at the  finite atom-light coupling strength some deviation from the fine tuning allows one to reduce significantly the width of the lowest energy band and also to have an almost perfect band geometry needed to simulate fractional Quantum Hall states. 
The
concluding Sec.~\ref{sec:Discussion} summarizes the findings.

\section{Hamiltonian}\label{sec:Hamiltonian}

We will consider a combined internal and center of mass atomic dynamics,
generally described by the Hamiltonian: 
\begin{equation}
\hat{H}=\frac{\mathbf{p}^{2}}{2M}+\hat{V}\,,\label{eq:H_full}
\end{equation}
where $\mathbf{p}=-i\hbar\mathbf{\nabla}$ is the momentum operator,
$M$ is the atomic mass, $\hat{V}\equiv\hat{V}\left(\mathbf{r}\right)$
is the position-dependent operator describing the atom-light coupling, and the hats are placed on operators affecting the atomic internal states. 

% [H]
\begin{figure}[tbh!]
\centering
\includegraphics[width=0.98\columnwidth]{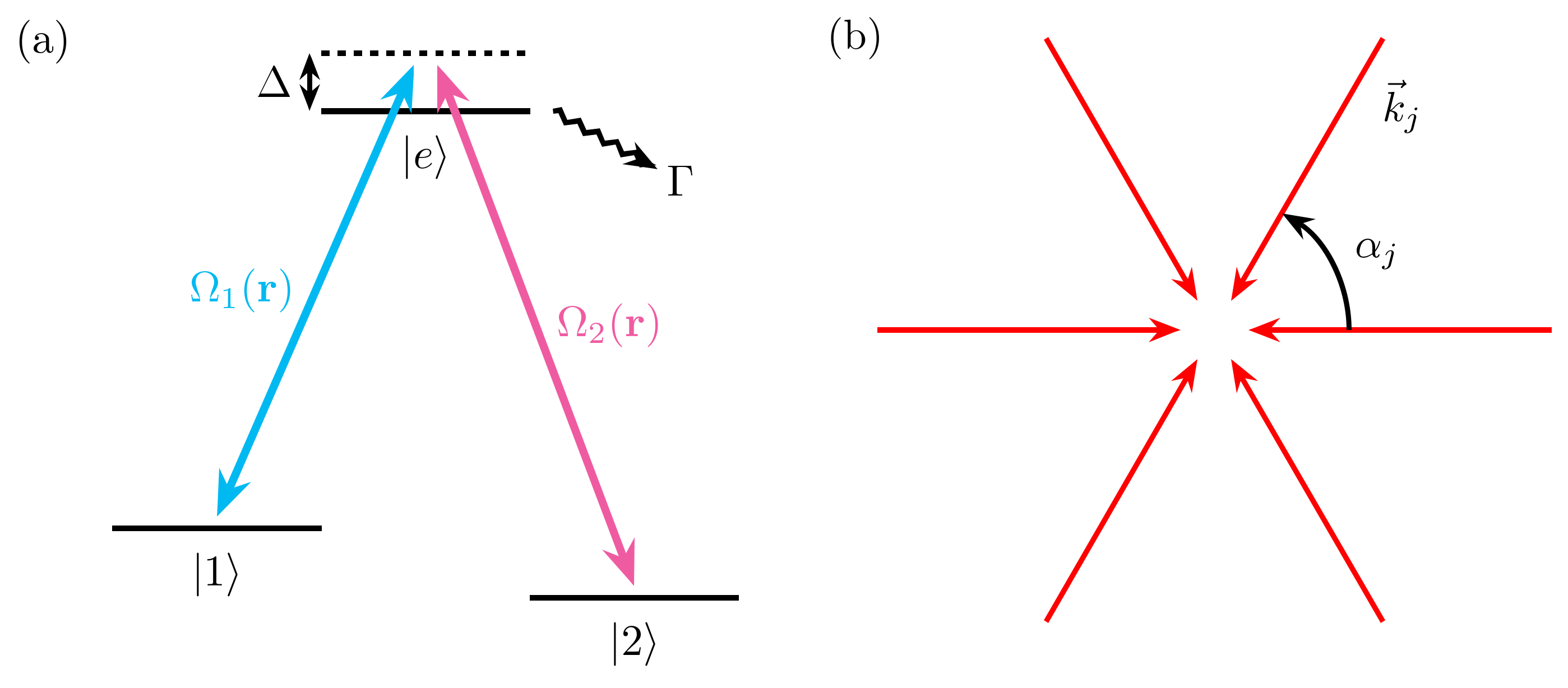}
\caption{
a) The Lambda scheme of the atom-light coupling. Laser fields characterized
by Rabi frequencies $\Omega_{1}$ and $\Omega_{2}$ couple resonantly
(or nearly resonantly with a single photon detuning $\Delta$) two
atomic ground states $\left|1\right\rangle $ and $\left|2\right\rangle $
with a common excited state $\left|e\right\rangle $. b) The AC condition
can be fulfilled if the Rabi frequency $\Omega_{1}$ represents a
superposition of the plane waves characterized by wave-vectors $\mathbf{k}_{j}$
with equal lengths $\left|\mathbf{k}_{j}\right|=\kappa$, whereas $\Omega_{2}$
is the corresponding superposition of the plane waves with wave-vectors
$-\mathbf{k}_{j}$, and properly chosen amplitudes and phases  (see
Sec.~\ref{sec:N-plane-waves}). }

\label{fig:Lambda} 
\end{figure}
We will consider the $\Lambda$ configuration of the atom-light coupling
depicted in Fig.~\ref{fig:Lambda}(a). The scheme involves two ground
states $\left|1\right\rangle $ and $\left|2\right\rangle $ which
are coupled resonantly (or nearly resonantly with a single photon
detuning $\Delta$) to the same excited state $\left|e\right\rangle $
by laser fields characterized by Rabi frequencies $\Omega_{1}$
and $\Omega_{2}$. As illustrated in Fig.~\ref{fig:Lambda}(b), the
Rabi frequencies $\Omega_{1}\equiv\Omega_{1}\left(\mathbf{r}\right)$
and $\Omega_{2}\equiv\Omega_{2}\left(\mathbf{r}\right)$ are generally
position dependent because of the position-dependence of the amplitude
or phase of the laser fields, making the atom-light coupling position
dependent. In what follows, we will mostly keep implicit this $\mathbf{r}$-dependence
of the Rabi frequencies and the atom-light coupling operator $\hat{V}\equiv\hat{V}\left(\mathbf{r}\right)$. 
Applying the rotating wave approximation \cite{Scully2008}, the atom-light
Hamiltonian reads for the $\Lambda$ scheme: 

\begin{equation}
\frac{\hat{V}}{\hbar}=-\tilde{\Delta}\left|e\right\rangle \left\langle e\right|+\frac{1}{2}\left[\left|e\right\rangle \left(\Omega_{1}\left\langle 1\right|+\Omega_{2}\left\langle 2\right|\right)+{\rm H.c.}\right]\,,\label{eq:V-definition}
\end{equation}
with $\tilde{\Delta}=\Delta+i\Gamma /2$, where an imaginary contribution $-i\Gamma/2$ has been added to the
energy of the excited state to represent its decay at the rate $\Gamma$.

The atom-light Hamiltonian (\ref{eq:V-definition}) can be cast in
terms of the excited and bright states 
\begin{equation}
\frac{\hat{V}}{\hbar}=\left(-\Delta-\frac{i}{2}\Gamma\right)\left|e\right\rangle \left\langle e\right|+\frac{\Omega}{2}\left(\left|e\right\rangle \left\langle B\right|+\left|B\right\rangle \left\langle e\right|\right)\,,\label{eq:V-definition-1}
\end{equation}
where the bright state $\left|B\right\rangle $ is a superposition
of the atomic ground states $\left|1\right\rangle $ and $\left|2\right\rangle $
directly coupled to the excited state $\left|e\right\rangle $: 
\begin{equation}
\left|B\right\rangle =\frac{1}{\Omega}\left(\Omega^{*}_{1}\left|1\right\rangle +\Omega^{*}_{2}\left|2\right\rangle \right)\,,\label{eq:B}
\end{equation}
and 
\begin{equation}
\Omega=\sqrt{\left|\Omega_{1}\right|^{2}+\left|\Omega_{2}\right|^{2}}\label{eq:Omega}
\end{equation}
is the total Rabi frequency.
Another superposition of the atomic ground
states orthogonal to the bright state is known as the dark state
\begin{equation}
\left|D\right\rangle =\frac{1}{\Omega}\left(\Omega_{2}\left|1\right\rangle -\Omega_{1}\left|2\right\rangle \right)\,.\label{eq:D}
\end{equation}

Introduction of the single photon detuning $\Delta$ in the atom-light coupling operator $\hat{V}$ allows one to eliminate the excited state $\left|e\right\rangle $ in the areas of a sufficiently small total Rabi frequency ($\Omega \ll \Delta$, with $\Delta \gtrsim \Gamma$), in particular, closer to the zero points of  $\Omega$ where the non-adiabatic transitons between the dark and bright states may be important.
Specifically, using the Schrieffer-Wolff transformation (outlined e.g. in
the Supplementary material of ref.~\cite{Yanes22PRL}), one arrives at the effective
Hamiltonian for the atomic dynamics in the ground state manifold comprised of the dark and bright states \cite{Nascimbene25PRL,Burba25PRR}
\begin{equation}
\hat{V}_{{\rm eff}}=V_{0}\left|B\right\rangle \left\langle B\right|\,,\label{eq:V-eff}
\end{equation}
 with
\begin{equation}
V_{0}=\frac{\hbar}{4}\frac{\Omega^{2}}{\Delta+i\Gamma/2}\,,\label{eq:V_0}
\end{equation}
 where the real and imaginary parts of $V_{0}$ describe, respectively,
the light shift and the decay of the bright state. Like in a number of previous treatments \cite{Zoller2016,Wang2018,Nascimbene25PRL,Burba25PRR}, we will concentrate on positive detuning $\Delta>0$. In that case, light shift is positive, so the energy of the lossy bright state is above that of the dark  state. This allows one to reduce the non-adiabatic transitions between the dark and bright states  taking place mostly in the spatial areas of small total Rabi frequency $\Omega \equiv \Omega\left(\mathbf{r}\right)$ (such as around its minima), where the condition $\Omega \ll \Delta$ holds, and thus it is relevant to use the projected description of the atom-light coupling in terms of the dark and bright states given by  Eqs.~\eqref{eq:V-eff}-\eqref{eq:V_0}.
Therefore, in the numerical simulations of the dark state spectra presented in Sec.~\ref{sec:Eigenstates-and-spectrum} we use a moderate positive detuning $\Delta$, as further discussed at the beginning of Sec.~\ref{sec:Eigenstates-and-spectrum}. 

It is noteworthy that the dark state is not featured in both the full and the  effective atom-light operators $\hat{V}$ or $\hat{V}_{{\rm eff}}$, so it
is an eigenstate of these operators with zero eigenenergy:
$\hat{V}|D\rangle=\hat{V}_{{\rm eff}}|D\rangle=0$. Note also that both the dark and the bright states are position-dependent
through the position-dependence of the Rabi frequencies $\Omega_{1}\left(\mathbf{r}\right)$
and $\Omega_{2}\left(\mathbf{r}\right)$.  This leads to the geometric potentials for atoms adiabatically following the dark state considered in the next Section~\ref{sec:Adiabatic-motion-in}. On the other hand, all numerical calculations presented in Sec.~\ref{sec:Eigenstates-and-spectrum} have been carried out using the full description in terms of Hamiltonian $\hat H$ with the atom-light coupling operator $\hat V$ that involves all three atomic internal states, i.e. the numerics do not rely on the adiabatic approximation or the effective atom-light coupling $\hat{V}_{{\rm eff}}$ given by \eqref{eq:V-eff}.

The dark and bright states can be defined up to a (position-dependent)
phase factor. The use of a different dark state corresponds to another
gauge choice for atoms adiabatically following the dark state. For
example, it is common to use the dark state \cite{Juz05PRA,Juz05JPB,Burba25PRR}
\begin{equation}
|\tilde{D}\rangle \equiv\frac{\left|\Omega_{2}\right|}{\Omega_{2}}\left|D\right\rangle =\frac{\left|1\right\rangle -\zeta\left|2\right\rangle }{\sqrt{1+|\zeta|^{2}}}\,,\label{eq:D-alt}
\end{equation}
which are expressed in terms of a relative Rabi frequency: 
\begin{equation}
\zeta=\Omega_{1}/\Omega_{2}=\left|\zeta\right|e^{iS},\label{eq:zeta}
\end{equation}
where $S$ is the corresponding relative phase. The two dark states
$\left|D\right\rangle $ and $|\tilde{D}\rangle $ differ
by the phase of the Rabi frequency $\Omega_{2}$. In our previous
work on 2D subwave-length lattices \cite{Gvozdiovas23PRA,Burba25PRR},
$\Omega_{2}$ was considered to be real, so the two sets of dark states
are equivalent. However, if $\Omega_{2}$ is a complex function and has
phase singularities, the dark state $|\tilde{D}\rangle $
might not be single valued at the singular points of $\Omega_{2}$
even if $\Omega\ne0$ there. On the other hand, the original dark
state $\left|D\right\rangle $ is single valued as long as the total
Rabi frequency $\Omega$ is non zero. In the present paper, we will
consider a situation where $\Omega_{2}$ is not necessarily real and
might have phase singularities. In that case, the use of the original
dark state $\left|D\right\rangle $ allows us to avoid the gauge-dependent
singularities of the geometric vector potential.

\section{Adiabatic approximation for the dark state atoms}\label{sec:Adiabatic-motion-in}

\subsection{Adiabatic Hamiltonian}

If the total Rabi frequency $\Omega$ greatly exceeds the characteristic
energy of the atomic center of mass motion, the atoms adiabatically
follow the lossless dark state. In that case the motion of the atomic center of
mass can be described by the dark-state wave-function
$\psi_{D}\left(\mathbf{r}\right)$ , the evolution of which is governed
by the Hamiltonian projected onto the dark-state manifold \cite{Juz05PRA,Juz05JPB,Dalibard11RMP,Goldman2014}:

\begin{equation}
\hat{H}_{{\rm D}}=\frac{1}{2M}\left(-i\hbar\mathbf{\nabla}-\mathbf{A}\right)^{2}+\phi\,,\label{eq:H_D}
\end{equation}
where $\mathbf{A}\equiv\mathbf{A}(\mathbf{r})$ and $\phi\equiv\phi(\mathbf{r})$
are, respectively, geometric vector and scalar potentials \footnote{In the condensed matter literature the geometric scalar potential $\phi$
is sometimes labeled by the letter $D$ \cite{Shi24PRB}}: 
\begin{equation}
\mathbf{A}\equiv\mathbf{A}_{DD}=i\hbar\left\langle D\right|\mathbf{\nabla}\left|D\right\rangle \quad\mathrm{and}\quad\phi=\frac{1}{2M}\mathbf{A}_{{\rm DB}}\cdot\mathbf{A}_{{\rm BD}}\,,\label{eq:A.phi}
\end{equation}
with
\begin{equation}
\mathbf{A}_{{\rm DB}}=i\hbar\left\langle D\right|\mathbf{\nabla}\left|B\right\rangle \quad\mathrm{and}\quad\mathbf{A}_{{\rm BD}}=\mathbf{A}^{*}_{{\rm DB}}\label{eq:A_BD}
\end{equation}
being the off-diagonal matrix elements of the vector potential that are responsible
for the non-adiabatic transitions between the dark and bright states.
The Hamiltonian (\ref{eq:H_D}) does not contain any additional position-dependent
light shift, as the dark state is the zero energy eigenstate of the
atom-light coupling operator. This facilitates simulation of the integer
and fractional Quantum Hall physics using ultracold atoms in the dark states.

\subsection{Vector and scalar potentials}

In the original gauge, the geometric vector potential corresponding
to the dark state $\left|D\right\rangle $ is: 

\begin{equation}
\begin{aligned}
\mathbf{A}&=\frac{i\hbar}{2\Omega^{2}}\sum^{2}_{j=1}\left(\Omega^{*}_{j}\boldsymbol{\nabla}\Omega_{j}-\Omega_{j}\boldsymbol{\nabla}\Omega^{*}_{j}\right)= \\
&=-\hbar\frac{|\Omega_{1}|^{2}\boldsymbol{\nabla}S_{1}+|\Omega_{2}|^{2}\boldsymbol{\nabla}S_{2}}{\Omega^{2}}\,,    
\end{aligned}
\label{eq:A_D-result}
\end{equation}
where $S_{1,2}$ are the phases of individual Rabi frequencies 
\begin{equation}
\Omega_{j}=|\Omega_{j}|e^{iS_{j}}\,,\quad j=1,2\, .\label{eq:Omega_p-amplitude. phase}
\end{equation}
On the other hand, in the transformed gauge, the geometric vector
potential $\mathbf{\tilde{A}}=i\hbar\langle \tilde{D}|\mathbf{\nabla}|\tilde{D}\rangle $
reads \cite{Juz05PRA,Juz05JPB} 
\begin{equation}
\mathbf{\tilde{A}}=\frac{i\hbar}{2}\frac{\zeta^{*}\mathbf{\nabla}\zeta-\zeta\mathbf{\nabla}\zeta^{*}}{1+|\zeta|^{2}}=-\hbar\frac{|\zeta|^{2}}{1+|\zeta|^{2}}\mathbf{\nabla}S\,,\label{eq:A_D-result-tilde}
\end{equation}
where $\zeta$ is the relative Rabi frequency introduced in Eq.~\eqref{eq:zeta}, $S$ being the corresponding relative
phase.

The geometric scalar potential $\phi$ and the magnetic field $\mathbf{B}=\boldsymbol{\nabla}\times\mathbf{A}=\boldsymbol{\nabla}\times\mathbf{\mathbf{\tilde{A}}}$
are gauge-independent, so they can be expressed in the same way for
both gauges as \cite{Juz05PRA,Juz05JPB}: 
\begin{equation}
\phi=\frac{\hbar^{2}}{2M}\mathbf{u}^{*}\cdot\mathbf{u},\quad\mathrm{and}\quad\mathbf{B}=i\hbar\mathbf{u}^{*}\times\mathbf{u}\,,\label{eq:phi-B-through u}
\end{equation}
where the vector $\mathbf{u}$ is proportional to the off diagonal matrix element of the vector potential $\mathbf{A}_{\mathrm{BD}}$ and reads

\begin{equation}
\mathbf{u}=\frac{\mathbf{\nabla}\zeta}{1+|\zeta|^{2}}=\frac{\Omega_{2}\mathbf{\nabla}\Omega_{1}-\Omega_{1}\mathbf{\nabla}\Omega_{2}}{\Omega^{2}}\,.\label{eq:u-vector}
\end{equation}
In what follows, we will be dealing with the atomic motion in the $xy$
plane, where the Rabi frequencies $\Omega_{1,2}$ and hence the vector
potential $\mathbf{A}$ depend only on the $x$ and $y$ coordinates,
 so the magnetic field $\mathbf{B}$ is perpendicular to the $xy$
plane.

\subsection{Magnetic flux for spatially periodic Rabi frequencies}\label{subsec:Magnetic-flux-for-spatially periodic Omegas}

In Sec.~\ref{sec:Symmetrically distributed plane waves} we will
consider a situation where the Rabi frequencies $\Omega_{1}$ and
$\Omega_{2}$ (and hence the atom-light coupling operator $\hat{V}(\mathbf{r})$)
are periodic in the $xy$ plane. Therefore, following refs.~\cite{Juz-Spielm2012NJP,Nascimbene25PRL,Burba25PRR}
let us present the general properties of the vector potential $\mathbf{A}=\mathbf{A}(\mathbf{r})$
and the corresponding magnetic field $\mathbf{B}=\boldsymbol{\nabla}\times\mathbf{A}$
for atoms adiabatically following an eigenstate of the spatial periodic
atom-light coupling operator $\hat{V}(\mathbf{r})$. In the present
paper such a dressed state is the dark state $\left|D\right\rangle =\left|D\left(\mathbf{r}\right)\right\rangle $,
but the same arguments hold for atoms adiabatically following an arbitrary
non-degenerate dressed state corresponding to any spatially periodic
atom-light coupling operator $\hat{V}(\mathbf{r})$ \cite{Juz-Spielm2012NJP}.
The dressed state and the corresponding vector potential are then
spatially periodic,  so the total magnetic flux over an elementary
cell is zero: 
\begin{align}
\alpha & =\oint_{{\rm cell}}\!\mathbf{A}\cdot d\mathbf{r}=0\,.\label{eq:alpha-flux}
\end{align}
Calling on the Stokes theorem, the total magnetic flux $\alpha$ can
be separated into two parts: 
\begin{align}
\alpha & =\alpha^{\prime}+\sum\oint_{{\rm singul}}\!\mathbf{A}\cdot d\mathbf{r}\,.\label{eq:alpha-flux-1}
\end{align}
The first term in Eq.~(\ref{eq:alpha-flux-1}) is the flux due to
the actual (continuous) magnetic field $\mathbf{B}=\nabla\times\mathbf{A}$
\begin{equation}
\alpha^{\prime}=\iint_{{\rm cell}}\!\mathbf{B}\cdot d\mathbf{S}\,.\label{eq:alpha-prime-flux}
\end{equation}
On the other hand, the second term in Eq.~(\ref{eq:alpha-flux-1})
involves integration around all (generally gauge-dependent) singular
points of the vector potential. Combining Eqs.~(\ref{eq:alpha-flux})-(\ref{eq:alpha-flux-1}),
the actual magnetic flux $\alpha^{\prime}$ over the elementary cell
is determined by the singularities of the vector potential within
the elementary cell \cite{Juz-Spielm2012NJP,Nascimbene25PRL,Burba25PRR}:
\begin{equation}
\alpha^{\prime}=-\sum\oint_{{\rm singul}}\!\mathbf{A}\cdot d\mathbf{r}\,.\label{eq:alpha-prime-flux-general-result}
\end{equation}

For the atom-light coupling involving two atomic internal states,
 the atomic dressed states can be parameterized by the azimuthal
and polar angles of the Bloch sphere. Such dressed states are not
single-valued in the south or north poles of the Bloch sphere, at
which the corresponding vector potential  can be singular \cite{Dalibard11RMP}.
By properly choosing the atom-light coupling, one can thus arrive
at a non-zero magnetic flux over a plaquette $\alpha^{\prime}$ equal
to an integer number of the Dirac quanta $2\pi\hbar$ \cite{Cooper11PRL,Juz-Spielm2012NJP}.

The situation is different for the $\Lambda$ scheme  in which the
dark state $\left|D\right\rangle $ given by Eq.~(\ref{eq:D}) is
fully determined by the Rabi frequencies $\Omega_{1}$ and $\Omega_{2}$,
so $\left|D\right\rangle $ is single valued, as long as the total
Rabi frequency $\Omega$ is non-zero.   Therefore, the adiabatic
motion of the dark state atoms is affected by the vector potential
which has no singularities, and thus the magnetic flux over a plaquette
is zero: $\alpha^{\prime}=0.$ The same holds for the dark state $|\tilde{D}\rangle$
corresponding to another gauge, Eq.~(\ref{eq:D-alt}), as the gauge
transformation cannot change the measurable magnetic flux $\alpha^{\prime}.$

Thus, one cannot create a non-staggered magnetic flux for the dark
state atoms of the $\Lambda$ scheme within the strict adiabatic approach.
Yet, the magnetic field can be composed of a smooth background magnetic
field and a set of narrow tubes of a strong magnetic field with an
opposite sign compensating the background flux \cite{Burba25PRR}.
If the flux tubes are narrow enough, they will not significantly influence
the atomic motion in the smooth background magnetic flux.  In the
limit where the width of the flux tubes goes to zero, only the non-staggered
background magnetic flux influences the atomic motion. Although
in that case the total Rabi frequency $\Omega$ goes to zero at the positions
of the infinitely narrow flux tubes, the specially chosen spatial
dependence of the Rabi frequencies around their zero points ensures
that the non-adiabatic transitions do not lead to any significant
losses there \cite{Nascimbene25PRL,Burba25PRR}. 

Note that even for infinitely narrow flux tubes the background
magnetic field is not perfectly uniform, which is different from the
 Landau problem for a charged particle in a uniform magnetic field
\cite{Nascimbene25PRL,Burba25PRR}. Recently, it was noticed that by
choosing the Rabi frequencies of the $\Lambda$ scheme to be a
superposition of three  plane waves with equal amplitudes \cite{Sommer-Cooper2509.01481},
the magnetic field can be non-staggered and obey the
Aharonov-Casher (AC) condition \cite{Aharonov-Casher79PRA}, leading
to the degeneracy of the lowest Landau level in the adiabatic limit. In the next Section, we will
present a general procedure of obtaining Rabi frequencies corresponding
to the AC conditions for  atoms adiabatically following
the dark state. Subsequently, we will turn to the analysis of the specific
situations. We will not restrict ourselves to the situation where the magnetic field is completely non-staggered \cite{Nascimbene25PRL,Sommer-Cooper2509.01481} and consider also effects due to deviation from the non-staggered magnetic field where the narrow patches of the magnetic field of the opposite sign are formed. In particular, it is shown that when the non-adiabatic corrections are included, such a slightly staggered magnetic field might be more favorable for making the lowest dark state band considerably narrower compared to that for the completely non-staggered magnetic field.   

\section{Perfect Chern insulators}\label{sec:Perfect-Chern-insulators}

\subsection{Another representation of the adiabatic Hamiltonian}

As we consider the atomic motion in the $xy$ plane perpendicular to
the magnetic field $\mathbf{B}=B_{z}\mathbf{e}_{z}$,   it is convenient
to introduce the operators \cite{Aharonov-Casher79PRA,Shi24PRB} 
\begin{equation}
\Pi_{\pm}=\Pi_{x}\pm i\Pi_{y}\,,\quad\mathrm{with}\quad\Pi_{x,y}=-i\hbar\partial_{x,y}-A_{x,y}\,.\label{eq:Pi_pm}
\end{equation}
In terms of these operators, the Hamiltonian \eqref{eq:H_D}  can
then be represented as 
\begin{equation}
\hat{H}_{{\rm D}}=\frac{1}{2M}\Pi_{\mp}\Pi_{\pm}+\phi\pm\frac{\hbar B_{z}}{2M}\,,\label{eq:H_D-alternative}
\end{equation}
where
\begin{equation}
B_{z}=\frac{1}{i\hbar}\left[\Pi_{x},\Pi_{y}\right]=\mathbf{e}_{z}\cdot\left(\boldsymbol{\nabla}\times\mathbf{A}\right)\,,\label{eq:Pi_u}
\end{equation}
is the magnetic field along the $z$ axis. The sign $\pm$ in Eqs.~\eqref{eq:Pi_pm}
and \eqref{eq:H_D-alternative} corresponds to the clockwise or anticlockwise
cyclotron motion in a non-staggered magnetic field with
\begin{equation}
\pm B_{z}\ge 0\,.\label{eq:B_z>0}
\end{equation}

\subsection{Aharonov-Casher condition}

 For the non-staggered magnetic field containing two or more Dirac flux quanta, the degenerate ground state appears if the last two terms in the Hamiltonian \eqref{eq:H_D-alternative} together make a constant $E_{\kappa}$
\begin{equation}
\phi\pm\frac{\hbar B_{z}}{2M}=E_{\kappa}\,,\quad\mathrm{with}\quad E_{\kappa}=\frac{\hbar^{2}\kappa{}^{2}}{2M}\,,\label{eq:phi.B_z--condition}
\end{equation}
where $\kappa$ is a wave-number corresponding to the energy $E_{\kappa}$.
Equation (\ref{eq:phi.B_z--condition}) represents the Aharonov-Casher
(AC) condition \cite{Aharonov-Casher79PRA,Shi24PRB}. Taking
$E_{\kappa}=0$, the condition (\ref{eq:phi.B_z--condition}) reduces
to the original AC condition obeyed by electrons with spins pointing
along the magnetic field \cite{Aharonov-Casher79PRA}. In that case,
the scalar potential $\phi$ in Eqs.~(\ref{eq:H_D-alternative}) and (\ref{eq:phi.B_z--condition})
represents the negative Zeeman energy $\mp\hbar B_{z}/2m$, and the
sign $\pm$ signifies the spin direction with respect to the magnetic
field. For such electrons, the lowest Landau level is degenerate and
is characterized by zero energy $E_{\kappa}=0$ even if the magnetic
field $B_{z}$ is inhomogeneous \cite{Aharonov-Casher79PRA}.

On the other hand, for atoms that adiabatically follow the dark state, condition (\ref{eq:phi.B_z--condition}) can hold only if the energy $E_{\kappa}$
is positive. In fact, the geometric scalar potential $\phi\propto\left|\mathbf{A}_{{\rm DB}}\right|^{2}$
is associated with the kinetic energy due to changes in the dark
state and therefore cannot be negative, so condition (\ref{eq:phi.B_z--condition})
can only hold for $E_{\kappa}>0$.  The non-zero $E_{\kappa}$ does
not change the physical situation, only shifts upwards the origin
of energy. Therefore, similar to the original AC problem \cite{Aharonov-Casher79PRA},
using the AC condition (\ref{eq:phi.B_z--condition}) and the additional
requirement (\ref{eq:B_z>0}), the eigenstates of
the operator $\Pi_{\pm}$ with zero eigenvalues form a set of degenerate ground states
of the Hamiltonian $\hat{H}_{{\rm D}}$ representing the lowest Landau
level with energy $E_{\kappa}$.

For the dark state atoms of the $\Lambda$ scheme, the general AC
condition (\ref{eq:phi.B_z--condition}) takes the form using Eq.~(\ref{eq:phi-B-through u})
for the geometric scalar potential $\phi$ and magnetic field
 $\mathbf{B}=B_{z}\mathbf{e}_{z}$:
\begin{equation}
\mathbf{u}^{*}\cdot\mathbf{u}\pm i\left(\mathbf{u}^{*}\times\mathbf{u}\right)\cdot\mathbf{e}_{z}=\kappa^{2}\,,\label{eq:u-condition}
\end{equation}
where $\mathbf{u}$ is given by Eq.~(\ref{eq:u-vector}). As the
vector $\mathbf{u}=\mathbf{e}_{x}u_{x}+\mathbf{e}_{y}u_{y}$ lies
in the xy plane, the AC condition (\ref{eq:u-condition}) simplifies
to  
\begin{equation}
v^{*}_{\pm}v_{\pm}=\kappa^{2}\,,\quad\mathrm{with}\quad v_{\pm}=u_{x}\pm iu_{y}\,.\label{eq:v^2 condition}
\end{equation}
The AC condition (\ref{eq:v^2 condition})  is equivalent to: 
\begin{equation}
v_{\pm}=i\kappa e^{i\chi}\,,\label{eq:v-condition}
\end{equation}
where the phase $\chi$ is an arbitrary real function. 

Calling on Eq.~(\ref{eq:u-vector}) for $\mathbf{u}$, the function
$v_{\pm}=u_{x}\pm iu_{y}$ can be represented in terms of Rabi
frequencies 
\begin{equation}
v_{\pm}=\frac{\Omega_{2}\partial_{\pm}\Omega_{1}-\Omega_{1}\partial_{\pm}\Omega_{2}}{\Omega^{2}}\,,\label{eq:v_pm-1}
\end{equation}
where
\begin{equation}
\partial_{\pm}=\partial_{x}\pm i\partial_{y}\label{eq:partial_pm}
\end{equation}
is the Wirtinger derivative.

\subsection{Procedure of obtaining {\normalsize\textmd{$\Omega_{1}$ and $\Omega_{2}$
}}providing perfect Chern insulators{\normalsize\textmd{ }}}\label{subsec:Procedure-for-obtaining Rabi frequencies}

 The Rabi frequencies obeying the AC condition ~(\ref{eq:v-condition}) can be obtained by the following general procedure. 

\textbf{\uline{Step 1}}: One should choose a Rabi frequency $\Omega_{1}\left(\mathbf{r}\right)$.
In particular, we will consider a situation where $\Omega_{1}$ represents
a sum of plane waves propagating in the $xy$ plane.

\textbf{\uline{Step 2}}: One should choose some $\chi$ and define
$\Omega_{2}\left(\mathbf{r}\right)$ by a relation
\begin{equation}
\Omega^{*}_{2}\left(\mathbf{r}\right)=-i\kappa^{-1}e^{-i\chi}\partial_{\pm}\Omega_{1}\left(\mathbf{r}\right)\label{eq:Omega_2*}
\end{equation}

\textbf{\uline{Step 3}}: One should ensure that $\Omega_{2}\left(\mathbf{r}\right)$
obtained from Eq.~(\ref{eq:Omega_2*}) is related to $\Omega_{1}\left(\mathbf{r}\right)$
as:
\begin{equation}
\Omega^{*}_{1}\left(\mathbf{r}\right)=i\kappa^{-1}e^{-i\chi}\partial_{\pm}\Omega_{2}\left(\mathbf{r}\right)\label{eq:Omega_1*}
\end{equation}
 If this is the case, substituting Eqs.~(\ref{eq:Omega_2*})-(\ref{eq:Omega_1*})
into Eq.~(\ref{eq:v_pm-1}), one can see that the required condition
(\ref{eq:v-condition}) holds.

\textbf{\uline{Step }}4: Finally we should make sure that the magnetic
field obtained for such Rabi frequencies is non-staggered and corresponds
to the proper sign presented in Eq.~(\ref{eq:B_z>0}), i.e. $\pm B_{z}\ge 0$.
 This is a demanding requirement. Following the arguments of Sec.~\ref{subsec:Magnetic-flux-for-spatially periodic Omegas},
the dark-state atoms of the $\Lambda$ scheme can be affected by the
non-staggered magnetic field only for the fine tuning of atom-light coupling when the
total Rabi frequency $\Omega$ has zero points, and the vector potential
$\mathbf{A}$ is singular at these points \cite{Burba25PRR,Nascimbene25PRL}.
Any deviation from this perfect situation leads to formation of sharp
peaks of the magnetic field at the zero points of the total Rabi frequency $\Omega$
corresponding to the fine tuning \cite{Burba25PRR}. The sharp
peaks of the magnetic field have the opposite sign compared to the
background magnetic field, making the total magnetic flux zero, as
required \cite{Burba25PRR}. Thus, the magnetic field becomes staggered
and condition $\pm B_{z}\ge 0$
no longer holds even if the requirement (\ref{eq:v-condition})
for the perfect Chern insulator is satisfied. Yet, as we will see, for sufficiently small deviations
from the fine tuning of atom-light coupling, the situation does not change drastically,
and the lowest Landau level remains nearly degenerate. 

Returning to Eqs.~(\ref{eq:Omega_2*})-(\ref{eq:Omega_1*}), using
these equations one finds that both Rabi frequencies $\Omega_{1}\left(\mathbf{r}\right)$
and $\Omega_{2}\left(\mathbf{r}\right)$ obeying the AC condition
\eqref{eq:v-condition}-\eqref{eq:v_pm-1} should be eigenfunctions of a linear differential
operator $S_{\pm}=-\kappa^{-2}\mathrm{e}^{\mathrm{i}\chi}\partial_{\mp}\left(\mathrm{e}^{-\mathrm{i}\chi}\partial_{\pm}\right)$
\begin{equation}
\Omega_{1}\left(\mathbf{r}\right)=S_{\pm}\Omega_{1}\left(\mathbf{r}\right)\,,\qquad\Omega_{2}\left(\mathbf{r}\right)=S_{\pm}\Omega_{2}\left(\mathbf{r}\right)\,,\label{eq:Omega_1 Omega_2 Conditions}
\end{equation}
where $S_{\pm}$ can be represented as
\begin{equation}
S_{\pm}=-\kappa^{-2}\left[\partial_{\mp}\partial_{\pm}-\mathrm{i}(\partial_{\mp}\chi)\partial_{\pm}\right].\label{eq:S}
\end{equation}
Since the Rabi frequencies $\Omega_{1}\left(\mathbf{r}\right)$ and
$\Omega_{2}\left(\mathbf{r}\right)$ are related to each other by
Eqs.~(\ref{eq:Omega_2*})-(\ref{eq:Omega_1*}), both conditions in
Eq.~\eqref{eq:Omega_1 Omega_2 Conditions} can be satisfied only if
the phase $\chi$ is constant. Specifically, in what follows we will
take $\chi=0$ without loss of generality. For constant $\chi$, the operator
$S_{\pm}$ becomes position-independent and reads
\begin{equation}
S_{\pm}=-\kappa^{-2}\left(\partial_{x}^2+ \partial_{y}^2\right)\,.\label{eq:S for constant chi}
\end{equation}
Hence, solutions to Eqs.~\eqref{eq:Omega_1 Omega_2 Conditions} and \eqref{eq:S for constant chi} (to be considered next) are superpositions of plane waves characterized by wave-vectors $\mathbf{k}_{j}$ of the same length $\kappa$ and lying in the $xy$ plane. 

\subsection{N plane waves}
\label{sec:N-plane-waves}

\subsubsection{Rabi frequencies}

Following the arguments presented above, let us consider a situation
where the Rabi frequency $\Omega_{1}\left(\mathbf{r}\right)$ is a
superposition of $N$ plane waves ($j=0\,,1\,,\ldots\,,N-1$) with (real) amplitudes $\Omega_{0j}$ and phases $\beta_{j}\pm\alpha_{j}$
\begin{equation}
\Omega_{1}\left(\mathbf{r}\right)=\sum^{N-1}_{j=0}\Omega_{0j}e^{i\beta_{j}\pm i\alpha_{j}}e^{i\mathbf{k}_{j}\cdot\mathbf{r}}\,,\label{eq:Omega_1-plane waves}
\end{equation}
where $\beta_{j}$ is an arbitrary phase and $\alpha_{j}$ is an azymuthal
angle that defines the wave vectors $\mathbf{k}_{j}$ lying in the
$xy$ plane 
\begin{equation}
\mathbf{k}_{j}=\kappa\left(\mathbf{e}_{x}\cos\alpha_{j}+\mathbf{e}_{y}\sin\alpha_{j}\right)\,.\label{eq:k_j-definition}
\end{equation}
  
Using
\begin{equation}
\partial_{\pm}e^{i\mathbf{k}_{j}\cdot\mathbf{r}}=i\left(k_{x}\pm ik_{y}\right)e^{i\mathbf{k}_{j}\cdot\mathbf{r}}=i\kappa e^{i\mathbf{k}_{j}\cdot\mathbf{r}\pm i\alpha_{j}}\,,\label{eq:partial_mp exp}
\end{equation}
Eqs.~(\ref{eq:Omega_1-plane waves}) and (\ref{eq:Omega_2*}) with $\chi=0$ provide
the following Rabi frequency $\Omega_{2}\left(\mathbf{r}\right)$:
\begin{equation}
\Omega_{2}\left(\mathbf{r}\right)=\sum^{N-1}_{j=0}\Omega_{0j}e^{-i\beta_{j}\mp2i\alpha_{j}}e^{-i\mathbf{k}_{j}\cdot\mathbf{r}}\,.\label{eq:Omega_2-plane waves}
\end{equation}
One can check that the Rabi frequencies $\Omega_{1}\left(\mathbf{r}\right)$
and $\Omega_{2}\left(\mathbf{r}\right)$ given by Eqs.~(\ref{eq:Omega_1-plane waves})
and (\ref{eq:Omega_2-plane waves}) also obey the second condition
of the perfect Chern insulator (\ref{eq:Omega_1*}).
This ensures that the AC condition (\ref{eq:v-condition}) is met.
Additionally, the magnetic field should be non-staggered ($\pm B_{z}\ge 0$)
for the perfect Chern insulator. The latter requirement can be fulfilled
only for fine-tuned parameters of the system, as we shall see in the next Section.

\subsubsection{Black state}
\label{Black state}

Since the Rabi frequencies $\Omega_{1,2}$ given by Eqs.~(\ref{eq:Omega_1-plane waves}) 
and (\ref{eq:Omega_2-plane waves}), obey the conditions \eqref{eq:Omega_1 Omega_2 Conditions} and \eqref{eq:S for constant chi}, they are eigenfunctions of the kinetic energy operator:
\begin{equation}
T\Omega_{1,2}\left(\mathbf{r}\right)=E_{\kappa}\Omega_{1,2}\left(\mathbf{r}\right)\,,\quad T=\frac{\mathbf{p}^{2}}{2M}\,,
\label{eq:p^2 Omega_12}
\end{equation}
where $E_{\kappa}$ is the recoil energy defined in Eq.~\eqref{eq:phi.B_z--condition}. Hence, one can introduce a state-vector 
\begin{equation}
\left|\mathrm{Black}\right\rangle =\mathcal{N}\left(\Omega_{2}\left(\mathbf{r}\right)\left|1\right\rangle -\Omega_{1}\left(\mathbf{r}\right)\left|2\right\rangle \right)=\mathcal{N}\Omega\left|D\right\rangle\,,
\label{eq:Black}
\end{equation}
which is an eigenvector of the kinetic energy: $T\left|\mathrm{Black}\right\rangle=E_{\kappa} \left|\mathrm{Black}\right\rangle$, where $\mathcal{N}$ is a normalization factor. Using also $\hat{V}\left|\mathrm{Black}\right\rangle=0$, one finds that the state vector $\left|\mathrm{Black}\right\rangle$ is an exact eigen-vector of the full Hamiltonian $\hat{H}=T+\hat{V}$:
\begin{equation}
\hat{H}\left|\mathrm{Black}\right\rangle = E_{\kappa}\left|\mathrm{Black}\right\rangle\,.
\label{eq:Pin-eigenstate}
\end{equation}

The state vector of the form \eqref{eq:Black} is called the black state \cite{VisserNienhuis1998}, because it not only belongs  to the dark state manifold, but is also completely immune to the non-adiabatic coupling with the bright state. 
The lossless black state \eqref{eq:Black}  is formed for any Rabi frequencies  $\Omega_{1,2}$ obeying Eq.~\eqref{eq:p^2 Omega_12} \cite{VisserNienhuis1998}, it is not necessary to impose the stronger AC conditions (\ref{eq:Omega_2*})-(\ref{eq:Omega_1*}) between the Rabi frequencies $\Omega_{1}$ and $\Omega_{2}$ to have such a lossless state.
The black states (non-absorbing trapping states) have been used in velocity-selective coherent population trapping (VSCPT) \cite{Aspect1988}. The black states have also been recently considered in the analysis of ultracold atoms affected by a fine tuned non-staggered synthetic magnetic field \cite{Nascimbene25PRL}. In the present paper (Sec.~\ref{Eluciation}) we make use of the black states in elucidating combined effects of the deviation from the fine tuning and the non-adiabatic coupling, which might lead to a significant narrowing of the lowest dark state band. 

\section{Plane waves with symmetrically distributed azymuthal angles }\label{sec:Symmetrically distributed plane waves}

In what follows, we will concentrate on a situation where the azymuthal
angles $\alpha_{j}$ characterizing the plane waves in Eqs.~(\ref{eq:Omega_2-plane waves})
and (\ref{eq:Omega_1-plane waves}) are symmetrically distributed
in the $xy$ plane:
\begin{equation}
\alpha_{j}=\frac{2\pi j}{N}\,,\quad j=0\,,1\,,\ldots\,,N-1\,.\label{eq:alpha_j_N-plane waves}
\end{equation}
In such a situation, not only the conditions for the perfect Chern insulator
(\ref{eq:Omega_2*})-(\ref{eq:Omega_1*}) are fulfilled, but also
the magnetic field can be non-staggered and have a proper sign ($\pm B_{z}\ge 0$) for three,
four and six plane waves ($N=3,\,4,\,6$). This ensures the remaining
condition (\ref{eq:B_z>0}) for the perfect Chern insulator. As
will be discussed in this Section, such a perfect situation appears
when all amplitudes $\Omega_{0j}$ and phases $\beta_{j}$ of
the constituting plane wave phases are equal: $\Omega_{0j}=\Omega_{0}$
and $\beta_{j}=0$ for all $j$. 

\subsection{Four plane waves $(N=4)$}\label{subsec:Four-plane-waves}

Let us start with a situation where the Rabi frequencies $\Omega_{1}$
and $\Omega_{2}$ are made of four symmetrically distributed
plane waves: $N=4$ and thus $\alpha_{j}=\pi j/2$. For generality, the amplitudes
of the counter-propagating waves are allowed to differ in the $x$
and $y$ directions and are equal to $\Omega_{0}/2$ and $b\Omega_{0}/2$,
respectively.   By  properly choosing the global phase of the Rabi
frequencies $\Omega_{1,2}$ and the origin of space,  the general
equations for the Rabi frequencies (\ref{eq:Omega_1-plane waves})
and (\ref{eq:Omega_2-plane waves}) take the form
\begin{equation}
\Omega_{1}\left(\mathbf{r}\right)=\Omega_{0}\left[\sin\left(\kappa x\right)\pm ibe^{-i\gamma}\sin\left(\kappa y\right)\right]\,,\label{eq:Omega_1-four-plane waves}
\end{equation}
\begin{equation}
\Omega_{2}\left(\mathbf{r}\right)=\Omega_{0}\left[\cos\left(\kappa x\right)-be^{i\gamma}\cos\left(\kappa y\right)\right]\,,\label{eq:Omega_2-four-plane waves}
\end{equation}
where $\gamma$ defines the relative phase between standing waves
in orthogonal directions $x$ and $y$. As we shall see, the remaining condition (\ref{eq:B_z>0})
for the perfect Chern insulator ($\pm B_{z}\ge 0$) holds if $b=1$ and
the phase $\gamma$ is equal to $\gamma=0,\,\pi/2\,,\pi,\,\mathrm{or}\,3\pi/2$
(modulo $2\pi$).
\begin{figure}[tbp]
\includegraphics[width=0.95\columnwidth]{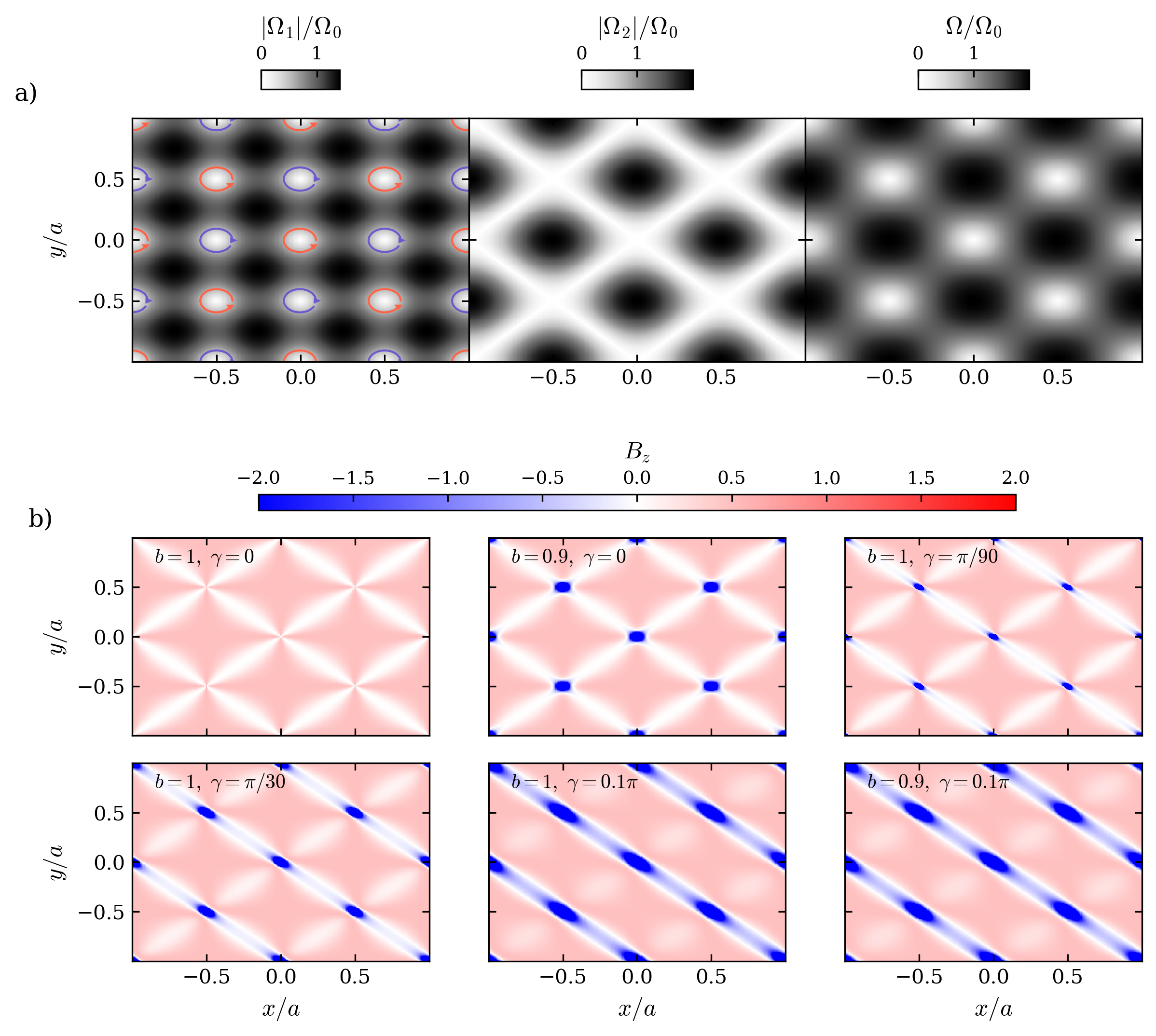}

\caption{(a) Rabi frequencies $\Omega_{1,2}$ given by Eqs.~(\ref{eq:Omega_1-four-plane waves})-(\ref{eq:Omega_2-four-plane waves})
together with the total Rabi frequency $\Omega$ for the fine tuned situation where $b=1$ and $\gamma=0$.
The arrows show the phase winding of $\Omega_{1}$ around its zero
points. (b) The corresponding effective magnetic field $B_{z}$ for $b$ and
$\gamma$ around $b=1$ and $\gamma=0$. Here and in other figures the distance is measured in the units of the lattice constant $a =2\pi/\kappa$. The geometric scalar potential
$\phi$ is  related to $B_{z}$ by the AC condition
(\ref{eq:phi.B_z--condition}), so $\phi$ is not displayed in this and subsequent plots of $B_{z}$.} 
\label{fig:magnetic-field-c-grid}
\end{figure}

Figure~\ref{fig:magnetic-field-c-grid}(a) presents
the Rabi frequencies given by Eqs.~(\ref{eq:Omega_1-four-plane waves})-(\ref{eq:Omega_2-four-plane waves})
with the upper sign of $\pm$ in $\Omega_{1}\left(\mathbf{r}\right)$
and $be^{i\gamma}=1$, while Fig.~\ref{fig:magnetic-field-c-grid}(b)
plots the corresponding effective magnetic field $B_{z}$ when $be^{i\gamma}$
is equal to unity or close to this value. The geometric scalar potential
$\phi$ is related to the effective magnetic field by the AC condition
(\ref{eq:phi.B_z--condition}), so it has not been plotted. For $be^{i\gamma}\ne1$
and $\left|be^{i\gamma}-1\right|\ll1$,   the magnetic field
is composed of a smooth positive background magnetic field and narrow
spots of a large negative magnetic field concentrated around the  minima
of the total Rabi frequency $\Omega$.  The latter narrow spots compensate
for the background magnetic field, so that the total magnetic flux over
an elementary cell is zero, as required by Eq.~(\ref{eq:alpha-flux}).
For \textbf{$be^{i\gamma}\rightarrow1$}, the narrow flux tubes become
infinitely narrow and reduce to Aharonov-Bohm singularities that contain
a unit flux quantum. Such singularities do not affect the atomic motion and are not
visible in the corresponding plot of Fig.~\ref{fig:magnetic-field-c-grid}(b).
Therefore, only the smooth positive magnetic field remains ($+B_{z}\ge 0$), so the
condition (\ref{eq:B_z>0}) for the perfect Chern insulator is obeyed. By taking the lower sign of $\pm$ in Eq.~(\ref{eq:Omega_1-four-plane waves})
for $\Omega_{1}\left(\mathbf{r}\right)$, the sign of the smooth magnetic
field is reversed and the condition (\ref{eq:B_z>0}) for the perfect
Chern insolator is obeyed again ($-B_{z}\ge 0$).  

The appearance of narrow magnetic flux tubes featured in Fig.~\ref{fig:magnetic-field-c-grid}
as $be^{i\gamma}$ is close to unity, can be understood from the
following arguments. The Rabi frequency $\Omega_{1}\left(\mathbf{r}\right)$
given by Eq.~(\ref{eq:Omega_1-four-plane waves}) goes to zero at
$\mathbf{r}=\mathbf{r}_{n,m}$, with
\begin{equation}
\mathbf{r}_{n,m}=\left(n\mathbf{e}_{x}+m\mathbf{e}_{y}\right)\pi/\kappa\,,\label{eq:r_n.m}
\end{equation}
 where $n$ and $m$ are integers. In a vicinity of these points
where $\mathbf{r}=\mathbf{r}_{n,m}+\mathbf{r}^{\prime}$ with $r^{\prime}\ll\pi/\kappa$,
the Rabi frequency $\Omega_{1}$ is linear in deviation $\mathbf{r}^{\prime}$,
while $\Omega_{2}$ is quadratic in $\mathbf{r}^{\prime}$. Thus,
the relative Rabi frequency $\zeta=\Omega_{1}/\Omega_{2}$ reads up
to the terms linear in the deviation $\mathbf{r}^{\prime}$:
\begin{equation}
\zeta=\kappa\frac{x^{\prime}\pm i\left(-1\right)^{m-n}be^{-i\gamma}y^{\prime}}{1-\left(-1\right)^{m-n}be^{i\gamma}}\,,\label{eq:zeta-around-r_n.m}
\end{equation}
For odd $m-n$ the denominator of Eq.~(\ref{eq:zeta-around-r_n.m})
is close to $2$, so the relative Rabi frequency $\zeta=\Omega_{1}/\Omega_{2}$
does not experience any steep changes. Hence, the effective magnetic
field and the geometric scalar potential are smooth in this area,
as one can see in Fig.~\ref{fig:magnetic-field-c-grid}(b). The situation
is quite different for even $m-n$ where the absolute value of the
denominator $1-be^{i\gamma}$ is much less than unity, so the
gradient of $\zeta$ is much larger than the wave-number $\kappa$ and hence a large magnetic flux surrounds $\mathbf{r}_{n,m}$.  A characteristic
radius $\rho_{0}$ over which the magnetic field is concentrated can
be estimated by taking $\left|\zeta\right|\approx1$, which gives the following:
\begin{equation}
\rho_{0}=\left|1-be^{i\gamma}\right|/\kappa\ll1/\kappa\,.\label{eq:rho_0}
\end{equation}
   For radial distances $r^{\prime}\equiv\rho$ a few times exceeding
$\rho_{0}$ , one has $|\zeta|^{2}\gg1$, so Eq.~(\ref{eq:A_D-result-tilde})
for the vector potential reduces to $\mathbf{\tilde{A}}\approx\mp\hbar\mathbf{\nabla}\varphi$,
with $\varphi$ being the azymuthal angle: $x^{\prime}=\rho\cos\varphi$
and $y^{\prime}=\rho\sin\varphi$. Therefore, for even $m-n$ the narrow
magnetic flux surrounding $\mathbf{r}_{n,m}$ is equal to $\mp2\pi\hbar$,
where $2\pi\hbar$ is the Dirac flux quantum. Within the elementary
cell, there are two such points corresponding to $n=m=0$ or $n=m=1$.
As pointed out in Sec.~\ref{subsec:Magnetic-flux-for-spatially periodic Omegas},
the total magnetic flux over the elementary cell is zero, so the background
magnetic flux over the elementary cell is $\pm4\pi\hbar$. The sign
$\pm$ is in agreement with the AC requirement \eqref{eq:B_z>0} saying
that $\pm B_{z}\ge 0$, as well as with the sign of the background magnetic field in Fig.~\ref{fig:magnetic-field-c-grid}(b). As $be^{i\gamma}\rightarrow1$, the width $\rho_{0}$
of the narrow magnetix fluxes goes to zero and only the smooth magnetic
field of proper sign ($\pm$) remains.  Following arguments
similar to those presented in ref.~\cite{Burba25PRR}, one can
show that there are no significant non-adiabatic losses due to
the narrow spots of the large magnetic field and the corresponding geometric scalar potential around the zero points of $\Omega_1$. 

\subsection{The cases where $N=3$ and $N=6$}\label{subsec:The-cases-where N eq 3 and 6}

For $N=3$ one can write $e^{\mp i4\pi j/N}=e^{\pm i2\pi j/N}$ in
the Rabi frequencies given by Eqs.~(\ref{eq:Omega_1-plane waves}),
(\ref{eq:Omega_2-plane waves}) and (\ref{eq:alpha_j_N-plane waves})
, so one has
\begin{equation}
\Omega_{1}\left(\mathbf{r}\right)=\sum^{3}_{j=1}\Omega_{0j}e^{\pm i2\pi j/3}e^{i\mathbf{k}_{j}\cdot\mathbf{r}}\,,\label{eq:Omega_1__three-plane waves}
\end{equation}
\begin{equation}
\Omega_{2}\left(\mathbf{r}\right)=\sum^{3}_{j=1}\Omega_{0j}e^{\pm i2\pi j/3}e^{-i\mathbf{k}_{j}\cdot\mathbf{r}}\,.\label{eq:Omega_2__three-plane waves}
\end{equation}
Here, all phases $\beta_{j}$ have been set to zero without
loss of generality by properly choosing the global phase of the Rabi frequencies
$\Omega_{1,2}$ and the origin of space.   This is not the case
if the Rabi frequencies $\Omega_{1,2}\left(\mathbf{r}\right)$ are made up of more than three plane waves.
For example, in the case of $N=4$ considered in the previous Section~\ref{subsec:Four-plane-waves},
one of the phases $\beta_{j}$ (or their combination) cannot be eliminated
by adjustment of the global phase of Rabi frequencies $\Omega_{1,2}$
and the origin of space. Therefore, the phase $\gamma$ is featured
in Eqs.~(\ref{eq:Omega_1-four-plane waves})-(\ref{eq:Omega_2-four-plane waves})
for $\Omega_{1,2}$. 

\begin{figure}[tbp]
\includegraphics[width=0.95\columnwidth]{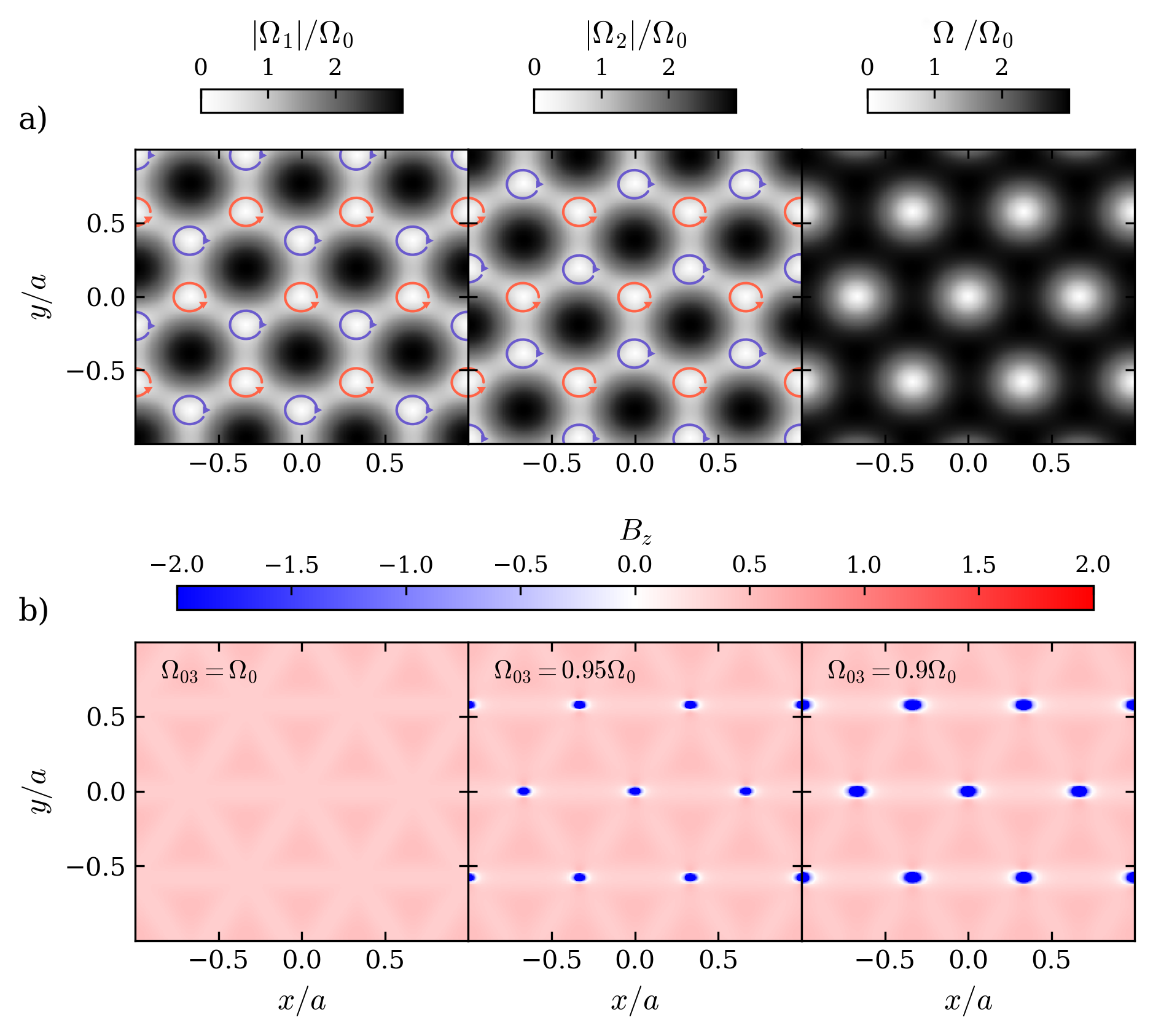}

\caption{(a) Rabi frequencies given by Eqs.~(\ref{eq:Omega_1__three-plane waves})-(\ref{eq:Omega_2__three-plane waves})
with $\Omega_{01}=\Omega_{02}=\Omega_{03}=\Omega_{0}$. The arrows
show the phase winding of $\Omega_{1}$ and $\Omega_{2}$ around their
zero points. (b) The corresponding effective magnetic field $B_{z}$ for $\Omega_{01}=\Omega_{02}=\Omega_{0}$
and the values of $\Omega_{03}$ close to $\Omega_{0}$.
}

\label{fig:omegas_bz_N3}
\end{figure}

In the fine tuned situation where all the amplitudes of the constituent plane waves are the same, $\Omega_{0j}=\Omega_{0}$ for all $j$, both Rabi frequencies  $\Omega_{1}$ and $\Omega_{2}$ go to zero
at $\mathbf{r}=0$, and also at all equivalent points of the triangular
lattice, as can be seen in Fig.~\ref{fig:omegas_bz_N3}(a). In a
vicinity of $\mathbf{r}=0$ the Rabi frequencies then read up to the
terms linear in the displacement $\rho$ 
\begin{equation}
\Omega_{1}\left(\mathbf{r}\right)=-\Omega_{2}\left(\mathbf{r}\right)=i\frac{3}{2}\Omega_{0}\kappa\left(x\pm iy\right)=i\frac{3}{2}\Omega_{0}\kappa\rho e^{\pm i\varphi}\,.\label{eq:Omega_1.2--three waves. approx}
\end{equation}
Using Eq.~(\ref{eq:A_D-result}) and (\ref{eq:Omega_1.2--three waves. approx}),
 the corresponding vector potential is $\mathbf{A}\approx\mp\hbar\mathbf{\nabla}\varphi$,
so there is an infinitely narrow magnetic flux $\mp2\pi\hbar$ concentrated
at $\mathbf{r}=0$. As within the elementary cell there is one such
point, following the arguments of Sec.~\ref{subsec:Magnetic-flux-for-spatially periodic Omegas},
the background magnetic flux over the elementary cell is $\pm2\pi\hbar$,
in agreement with the AC requirement \eqref{eq:B_z>0} that $\pm B_{z}\ge 0$.
This is also seen in the plots of the effective magnetic field presented in Fig.~\ref{fig:omegas_bz_N3}(b) for
the fine tuned situation where $\Omega_{0j}=\Omega_{0}$ for all $j$. 
\begin{figure}[tbp]
\includegraphics[width=0.95\columnwidth]{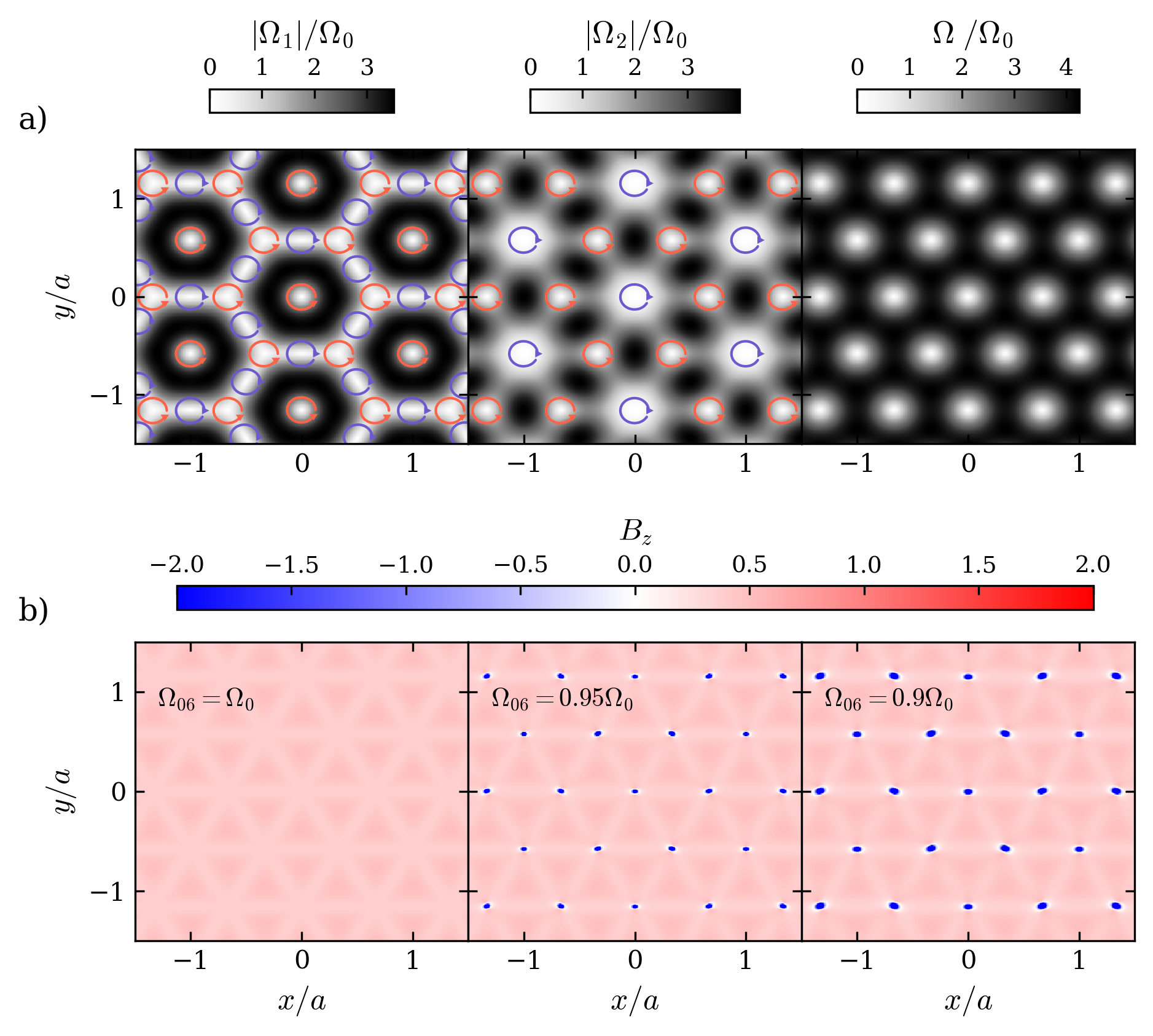}

\caption{(a) Rabi frequencies given by Eqs.~(\ref{eq:Omega_1-plane waves}),
(\ref{eq:Omega_2-plane waves}) and (\ref{eq:alpha_j_N-plane waves})
with $N=6$ for $\beta_{j}=0$ and $\Omega_{0j}=\Omega_{0}$, with
$j=1,\,2,\ldots\,,6$. The arrows show the phase winding of $\Omega_{1}$
and $\Omega_{2}$ around their zero points. (b) The corresponding
magnetic field $B_{z}$ when $\Omega_{0j}=\Omega_{0}$ for $j\protect\ne3$
and $\Omega_{03}$ is close to $\Omega_{0}$.}

\label{fig:omegas_bz_N6}
\end{figure}

Note that for three fine tuned plane waves ($N=3$) with $\Omega_{0j}=\Omega_{0}$ for all $j$, the magnetic field shown in the left plot of Fig.~\ref{fig:omegas_bz_N3}(b) is more homogeneous than the corresponding magnetic field for $N=4$ presented in the left plot of Fig.~\ref{fig:magnetic-field-c-grid}(b). This is because for $N=4$, the Rabi frequency $\Omega_2$ shown in Fig.~\ref{fig:magnetic-field-c-grid}(a)  is zero on the diagonals in the $xy$ plane, leading to the corresponding white patches around these diagonals for the magnetic field shown in the left plot of Fig.~\ref{fig:magnetic-field-c-grid}(b). This is not the case for $N=3$ where both Rabi frequencies $\Omega_1$ and $\Omega_2$ shown in Fig.~\ref{fig:omegas_bz_N3}(a) go to zero only at isolated points.
Note also that if the amplitudes $\Omega_{0j}$ of the constituent $N=3$ plane waves  are not exactly equal,
the narrow flux tubes of the sign opposite to that
of the background magnetic field appear around $\mathbf{r}=0$, as
well as around the equivalent points of the triangular lattice, as one can see in 
Fig.~\ref{fig:omegas_bz_N3}(b). The
situation is similar if one superimposes six plane waves ($N=6$),
as illustrated in Fig. \ref{fig:omegas_bz_N6}.

\subsection{The case where $N=5$}

\begin{figure}[t]
\includegraphics[width=0.95\columnwidth]{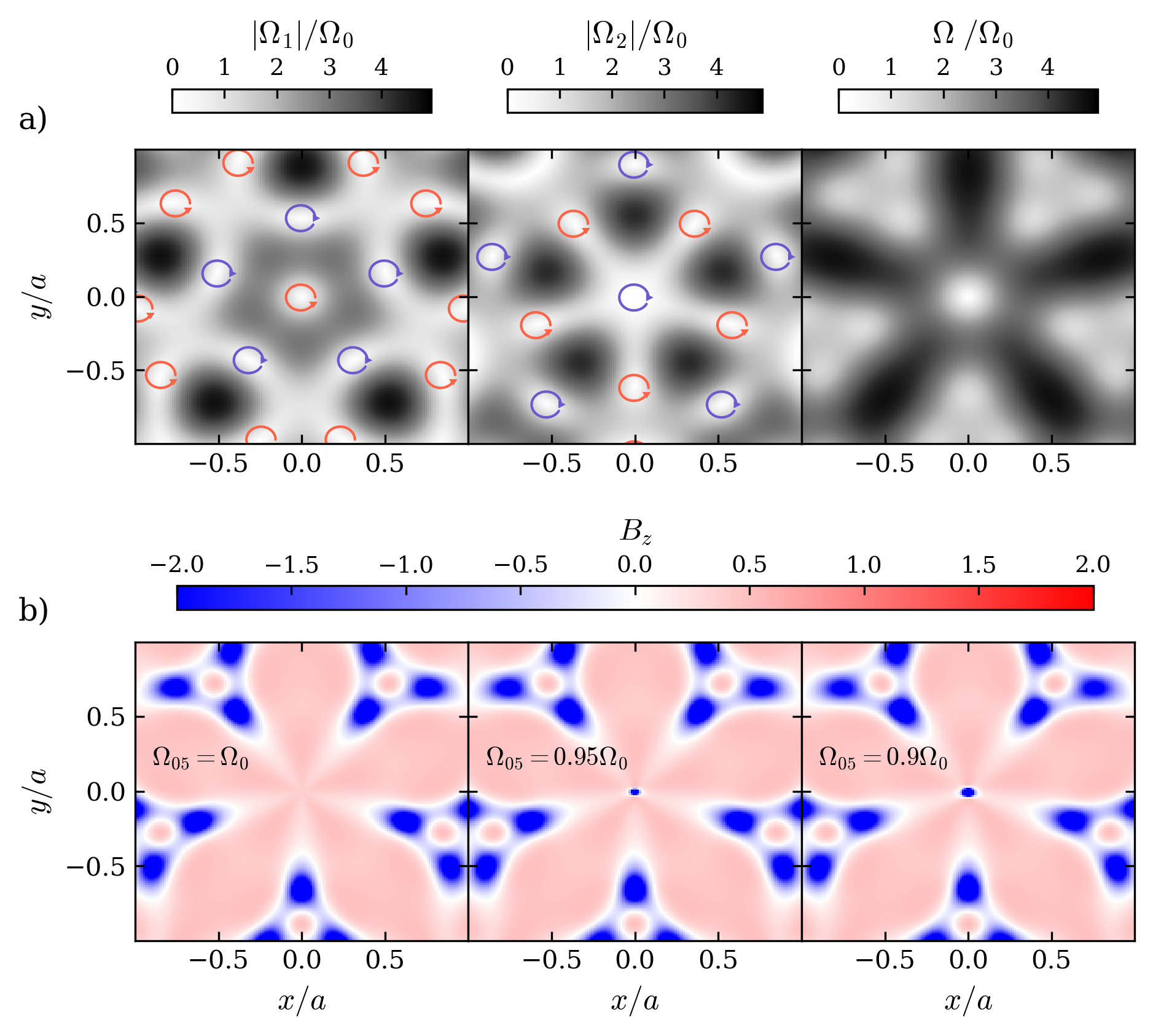}

\caption{(a) Rabi frequencies given by Eqs.~(\ref{eq:Omega_1-plane waves}),
(\ref{eq:Omega_2-plane waves}) and (\ref{eq:alpha_j_N-plane waves})
with $N=5$ for $\beta_{j}=0$ and $\Omega_{0j}=\Omega_{0}$, with
$j=1,\,2,\ldots\,,5$. The arrows show the phase winding of $\Omega_{1}$
and $\Omega_{2}$ around their zero points. (b) The corresponding
magnetic field $B_{z}$ for various values of $\Omega_{05}$ when $\Omega_{0j}=\Omega_{0}$
for $j\protect\ne5$. }

\label{fig:omegas_bz_N5}
\end{figure}

By superimposing five symmetrically distributed plane
waves, the corresponding Rabi frequencies $\Omega_{1}$ and $\Omega_{2}$ [given by Eqs.~(\ref{eq:Omega_1-plane waves}),
(\ref{eq:Omega_2-plane waves}) and (\ref{eq:alpha_j_N-plane waves})
with $N=5$] provide a magnetic field composed of a quite smooth background
magnetic flux and patches of the magnetic field of opposite sign, making
a quasicrystalline pattern shown in Fig. \ref{fig:omegas_bz_N5}. One
can see that the latter patches of the opposite magnetic field appear
even for fine tuning (Fig. \ref{fig:omegas_bz_N5}(a)), so one cannot create a non-staggered
quasicrystalline magnetic field. This is because, unlike for the periodic distribution of Rabi frequencies corresponding to $N=3,\,4,\,6$, in the case of $N=5$ there are no coinciding zero points of the Rabi frequencies $\Omega_{1}$
and $\Omega_{2}$ except for the coordinate origin $\mathbf{r}=0$, as can be seen when comparing Fig. \ref{fig:omegas_bz_N5} (a) with the previous plots of Figs. \ref{fig:magnetic-field-c-grid} -
\ref{fig:omegas_bz_N6}.
Some deviation from the fine tuning
does not change the situation much; only an additional narrow peak
of the opposite magnetic field appears at $\mathbf{r}=0$, as one
can see in Fig. \ref{fig:omegas_bz_N5} (b). We studied the spectrum for this quasicrystal structure using open boundary conditions. Unlike the fine tuned  periodic system involving $N=3$ or $N=4$ plane waves considered in the next Section \ref{sec:Eigenstates-and-spectrum}, for $N=5$ we did not find any chiral edge states, as the patches of the opposite magnetic field (blue areas in Fig. \ref{fig:omegas_bz_N5}) are too large. Note that similar findings were recently reported in ref.~\cite{Burgess26PRR}.   

\section{Eigenstates and spectrum}\label{sec:Eigenstates-and-spectrum}

\subsection{Chiral edge states}\label{sec:chiral-edge-states}
To study the topological properties of the system, we will commence with a cylindrical geometry for $N=3$ and $N=4$, in which the system is periodic along the $x$ direction and has a finite width $L$ along the $y$ direction, ranging from $y=-L/2$ to $y=L/2$.
(The bulk spectrum and the effects of losses will be considered later in Sec~\ref{Bulk spectrum}). In the periodic $x$ direction, Bloch’s theorem is applied and the eigenstates are labeled by the quasimomentum $q_{x}$. Along the transverse direction, Dirichlet boundary conditions are imposed, leading to a discrete set of modes. This configuration allows one to distinguish the bulk and edge states directly from the spectrum~\cite{Hatsugai1993,Niu2010RMP}.

\begin{figure}[tbp]
    \centering
    \includegraphics[width=0.9\linewidth]{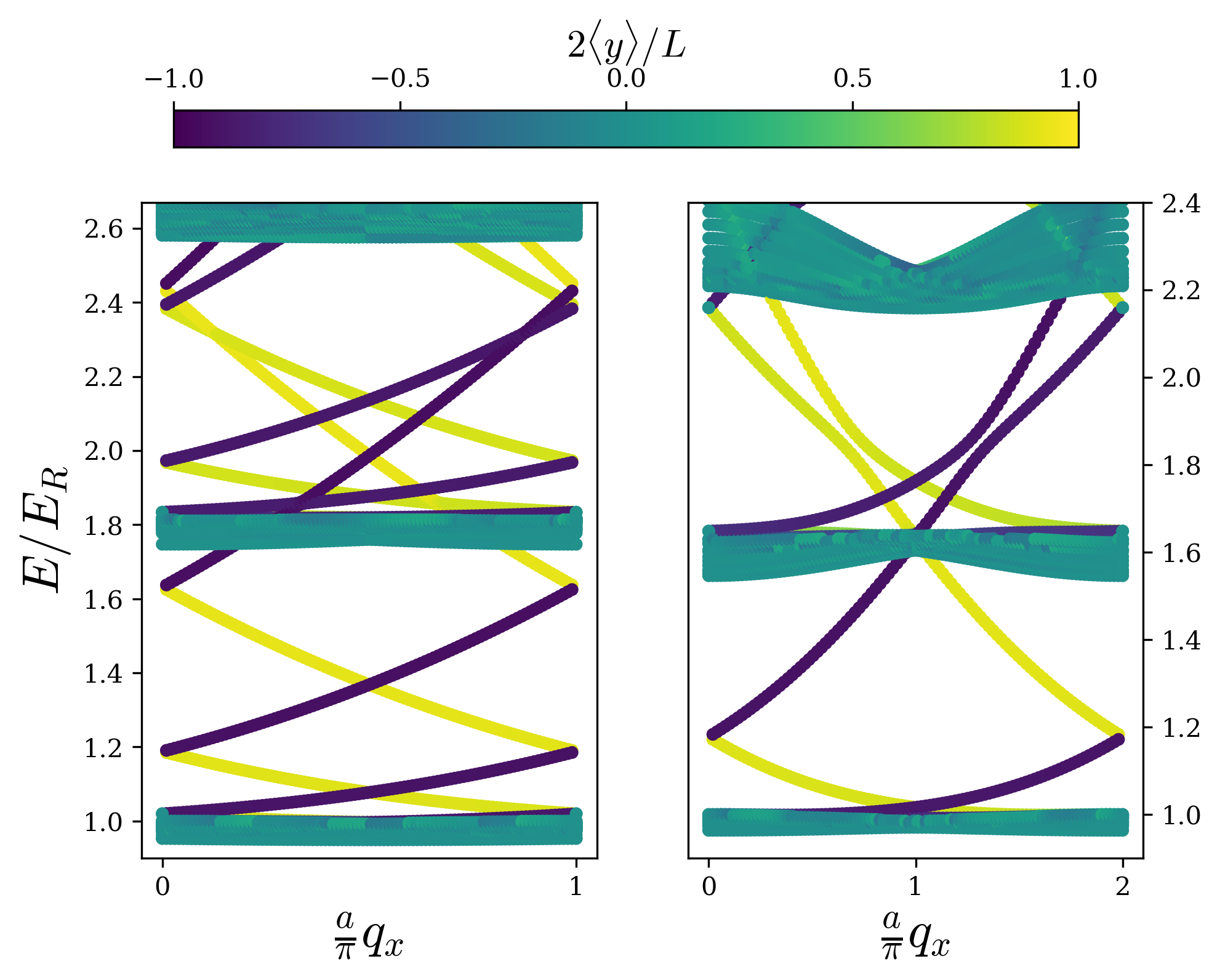}
    \caption{
    Dependence of eigenenergies $E$ on the quasimomentum $q_{x}$ in cylindrical geometry with the transverse width $L=8a$. Left and right panels correspond  the atom-light coupling involving three $(N=3)$ and four $(N=4)$ plane waves, respectively. Both panels represent ideal cases ($\Omega_{0j}=\Omega_{0}=4000E_{R}$ and $\beta_{j}=0$ for all $j$) with a moderate detuning ($\Delta = \Omega_0$) 
    and no losses ($\Gamma =0$). 
 Colors indicate the center of mass position $2\langle y \rangle/L$.
 In this and the subsequent figures, the calculations have been carried out using the full Hamiltonian $\hat H$ given by Eqs.\eqref{eq:H_full}-\eqref{eq:V-definition} with all three atomic internal states involved.
 } 
    \label{fig:N_3_4_cyl_spectr_panels}
\end{figure}

In Fig.~\ref{fig:N_3_4_cyl_spectr_panels} we plot the energy spectrum of the eigenstates as a function of quasimomentum $q_{x}$ for an ideal case where the Rabi frequencies given by Eqs. (\ref{eq:Omega_1-plane waves}) and (\ref{eq:Omega_2-plane waves}) have equal amplitudes ($\Omega_{0j}=\Omega_0$) and  phases ($\beta_{j}=0$) for all $j$, and there is a moderate detuning ($\Delta = \Omega_0$).  
The use of a positive detuning $\Delta>0$ comparable to the amplitude of Rabi frequencies ($\Delta \sim \Omega_0$) is common in theoretical and experimental studies on geometric potentials for dark state atoms \cite{Zoller2016,Wang2018,Nascimbene25PRL,Burba25PRR}. In that case, the detuning $\Delta$ is not too big, so  the concept of dark and bright states still works quite well \cite{McDonnell04PRA}, while, as discussed in the paragraph following Eq.~\eqref{eq:V_0}, moderate positive detuning  $\Delta$ reduces the losses.
Furthermore, in the numerical analysis of this Section \ref{sec:Eigenstates-and-spectrum}, the amplitude of Rabi frequencies $\Omega_0$ is several thousand recoil frequencies (as in the experiment \cite{Wang2018} with $^{171}\mathrm{Yb}$ atoms) or less. Note that the recoil frequency is equal to  $E_{\mathrm R}= E_{\kappa} \approx 2\pi\times  3.8 \times 10^3 \mathrm{Hz}$  for commonly used $^{171}\mathrm{Yb}$ \cite{Wang2018} or $^{87}\mathrm{Rb}$ \cite{Steck2003Rubidium87} atoms (from now on the energy is measured in frequency units by setting $\hbar=1$). 

In the case of $N=3$ (Fig. \ref{fig:N_3_4_cyl_spectr_panels}, left panel), calculations were performed using a supercell of length $2a$ in the $x$ direction. As one can see in Fig.~\ref{fig:omegas_bz_N3}(a), the actual spatial periodicity of the atom-light coupling is three times smaller and equals $\tilde a = 2a/3$ in the $x$ direction. Thus, the supercell contains three primitive unit cells of the optical potential, so the Brillouin zone associated with the supercell is three times smaller than the primitive Brillouin zone. As a result, the Bloch bands are
folded three times into the reduced Brillouin zone.
This leads to multiple band crossings and repeated structures in the dispersion, reflecting the enlarged real-space periodicity rather than additional physical degeneracies.
For $N=4$ corresponding to the geometry of the square lattice, we take the periodicity in the x direction to coincide with the lattice constant $a$. Yet, the actual periodicity of the atom-light coupling is twice as small, as in previous related studies \cite{Gvozdiovas23PRA,Burba25PRR}. Thus, the edge bands shown in the right panel of Fig.~\ref{fig:N_3_4_cyl_spectr_panels} are twice folded.

To identify the edge states, one can characterize each eigenstate by its (dimensionless) center of mass coordinate $2\langle y \rangle/L$.
Bulk states 
are centered at $\langle y \rangle \approx 0$, while the edge states are located near the boundaries with $2\langle y \rangle/L\approx 1$ or $2\langle y \rangle/L\approx -1$ depending on the boundary. In Fig.~\ref{fig:N_3_4_cyl_spectr_panels} we see that the edge states traverse the band gaps and connect adjacent bands, demonstrating nontrivial band topology. According to the bulk-edge correspondence \cite{Hatsugai1993,Niu2010RMP} the difference in the number of chiral edge modes above and below the given energy band determines the Chern number associated with that band. Thus, the bands shown in Fig.~\ref{fig:N_3_4_cyl_spectr_panels} are characterized by the unit Chern number, in agreement with the bulk calculations. 

Although the plots have been made for the perfect AC situation, the lowest band has a finite width due to the finite size of the system in the transverse (y) direction. 
Furthermore, in such a perfect AC situation, the lowest band has some finite residual width even for the periodic boundary conditions in both directions because of some deviation from the adiabatic approximation due to the finite Rabi frequencies used, as we shall see next. Note also that for $N=3$ the higher bands are narrower than for $N=4$. This is because the magnetic field is more uniform for $N=3$ than for $N=4$ (see Sect.~\ref{subsec:The-cases-where N eq 3 and 6}), so for $N=3$ the problem is closer to the Landau problem for the higher bands. Yet this does not apply to the lowest energy band, in which the AC condition between the magnetic field and the scalar potential should make the lowest band completely flat for the bulk system and the adiabatic limit even if the magnetic field is not uniform.

Note that Fig.~\ref{fig:N_3_4_cyl_spectr_panels}, as well as subsequent Figs.~\ref{fig:bulk_spectrum_N4_4panel}-\ref{fig:bandwidth-N4}, display the spectral region associated with the dark-state bands, rather than the global lowest eigenvalues of the full Hamiltonian. The lower excited-state-like branch [eliminated in the effective ground state Hamiltonian, Eqs.~\eqref{eq:V-eff}-\eqref{eq:V_0}] is situated far below this displayed region. Its absence from the plot does not mean that the excited state was omitted from the calculation: The results of numerical simulations presented in this and subsequent plots retain all three atomic internal states.

\subsection{Bulk spectrum and losses}
\label{Bulk spectrum}

\begin{figure}[tbp]
\includegraphics[width=1.0\columnwidth]{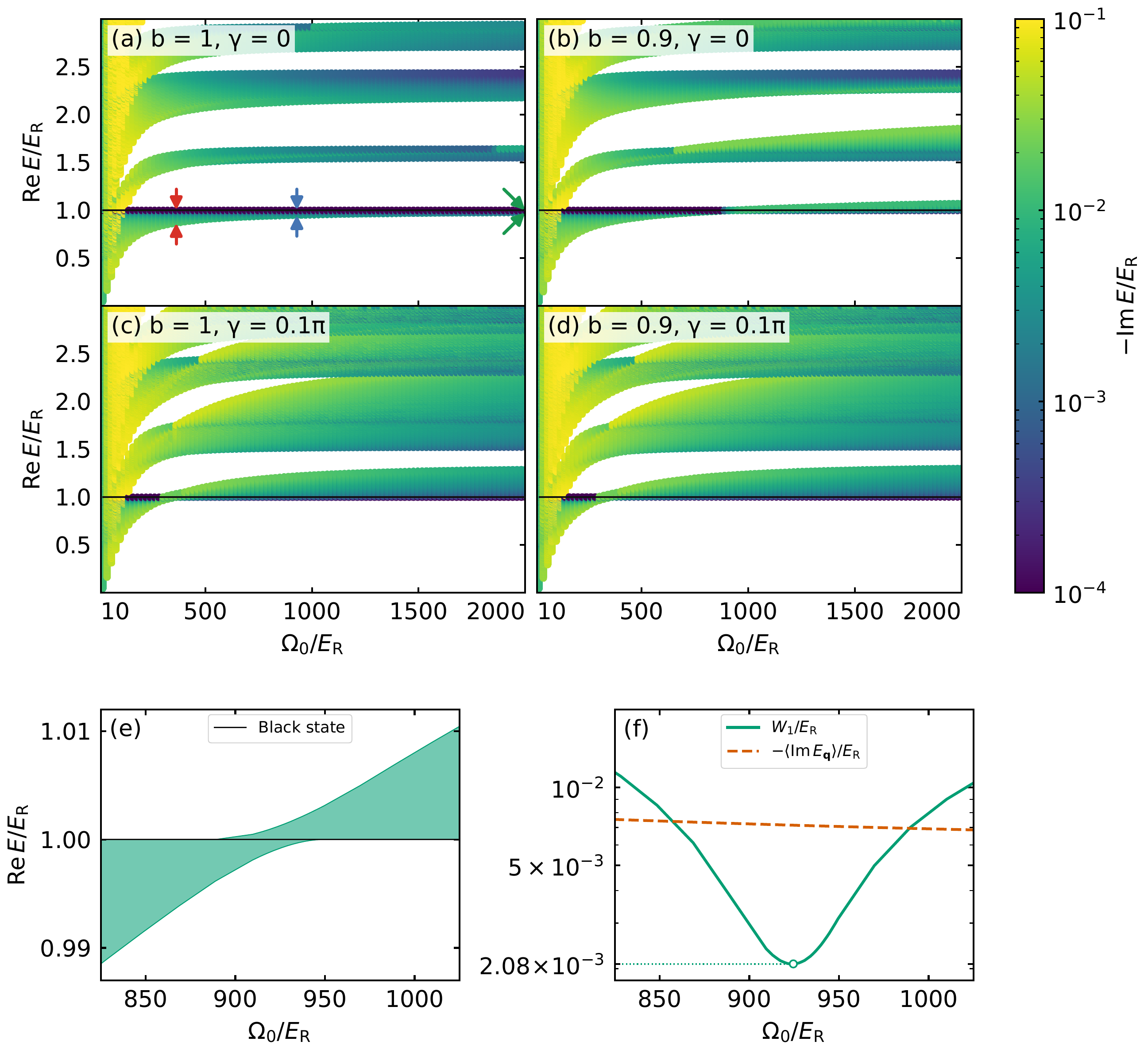}
\caption{
(a)--(d) Real parts of the eigenenergies, $\operatorname{Re}E$ versus the Rabi frequency $\Omega_0$ (in the units of recoil energy $E_{\mathrm R}$) for square geometry ($N=4$)
in the spectral region of the dark-state manifold.
The panels (a)--(d) use $(b,\gamma)=(1,0),(0.9,0),(1,0.1\pi),(0.9,0.1\pi)$,
respectively. (e) Enlarged lowest-band spectrum,
(f) its width $W_1/E_{\mathrm R}$ (solid green) and band averaged imaginary part of the energy 
$-\langle\operatorname{Im}E_{\mathbf{q}}\rangle/E_{\mathrm R}$ (dashed orange) for $(b,\gamma)=(0.9,0)$ corresponding to the panel (b). 
The red, blue and green arrows in (a) indicate the bandwidths of $0.13E_{\mathrm R}$, $0.051E_{\mathrm R}$ and $0,023 E_{\mathrm R}$  for relative Rabi frequencies $\Omega_0/E_{\mathrm R}$ equal to  $360,\, 925\,\, \mathrm{and} \,\, 2000$, respectively, the first two arrows corresponding to the minimum widths of the lowest band in the panels (c) and (b). In (a)-(e) colors show $-\operatorname{Im}E/E_{\mathrm R}$ on a common logarithmic
scale; the lossless black state at $E/E_{\mathrm R}=1$ is drawn in black above the colored spectrum. In all panels of this figure we take $\Delta=2000E_{\mathrm R}$ and $\Gamma = 1000E_{\mathrm R}$. The calculations have been carried out using $N_{\mathrm{F}}=21$ Fourier modes in both spatial directions $x$ and $y$, and the convergence of the spectra has been checked.
}
\label{fig:bulk_spectrum_N4_4panel}
\end{figure}

Let us now turn to the bulk energy spectrum of the Hamiltonian $\hat H$ by analyzing its dependence on the Rabi frequency $\Omega_0$ that defines the applicability of the adiabatic approximation used in deriving the Aharonov-Casher conditions for atoms adiabatically following the dark state.
We focus on the square geometry ($N=4$); results for the triangular configuration ($N=3$) are qualitatively similar. 

Figures~\ref{fig:bulk_spectrum_N4_4panel} (a)-(e) show a real part of the energies of the Bloch states $\mathrm{Re}E\equiv \mathrm{Re}E_{\mathbf{q}}$  vs. the Rabi frequency $\Omega_0$
 for detuning  $\Delta=2000E_{\mathrm R}$ and loss rate $\Gamma=1000E_{\mathrm R}$, the latter value corresponding to a  characteristic rate of excited state decay of alkali atoms, such $^{87}\mathrm{Rb}$ ($\Gamma/E_{\mathrm R}=1600$) or $^{39}\mathrm{K}$  ($\Gamma /E_{\mathrm R}=696$)  \cite{Metcalf1999LaserCooling,Steck2003Rubidium87}.  
In Figs.~\ref{fig:bulk_spectrum_N4_4panel} (a)-(e) the blue-green-yellow colors indicate an  imaginary part of the energies $\mathrm{Im} E_{\mathbf{q}}$ of the Bloch states, related to their decay rate $\Gamma_{E_{\mathbf{q}}}$ as:  
\begin{equation}
\Gamma_{E_{\mathbf{q}}} = n_{e\mathbf{q}} \Gamma=-2\mathrm{Im}E_{\mathbf{q}} /\hbar\,,
\label{eq:Gamma_E_q}
\end{equation}
where $n_{e\mathbf{q}}$ is a population of the lossy excited atomic internal state $\left|e\right\rangle$  for a Bloch state  characterized by a quasi-momentum $\mathbf{q}$, and we have already set $\hbar=1$. 
The excited state population $n_{e\mathbf{q}}$ is due to non-adiabatic corrections to the dark state bands, so $n_{e\mathbf{q}}$ is tiny %$n_{e\mathbf{q}}\ll1$
for sufficiently large values of $\Omega_0$, for which the adiabatic approach works quite well. For $\Omega_0$ exceeding some $200 E_{\mathrm R}$, the imaginary part of the eigenenergies of the lowest dark state band is in the green-blue region in Fig.~\ref{fig:bulk_spectrum_N4_4panel} (a)-(e), in which $-\mathrm{Im}E_{\mathbf{q}}/E_{\mathrm R}$ is around $10^{-2}$ to $10^{-3}$ or less; the corresponding excited state population $n_{e\mathbf{q}}$ is on the order of $10^{-5}$ or $10^{-6}$ or smaller. 
Note that the excited state population is determined mostly by non-adiabatic corrections to the dark state bands due to finite $\Omega_0$, it is not sensitive to the decay rate $\Gamma$. In fact, we have checked that almost the same values of $n_{e\mathbf{q}}$ are obtained (with a difference of only $0.1$ percent) by decreasing the decay rate of the excited state from $\Gamma = 1000E_{\mathrm R}$ to $\Gamma = 0$. Hence,  as $\Gamma$ decreases $l$ times, the value of $\operatorname{Im}{E_{\mathbf{q}}}$ also becomes $l$ times smaller. On the other hand, the real part of the eigen-energies $\operatorname{Re}E_{\mathbf{q}}$ is also found to be non-sensitive to the decay rate, and the difference between its values for $\Gamma = 1000E_{\mathrm R}$ and $\Gamma = 0$ is below $0.05$ percent. We will return to these issues later in the Section when discussing the modification of the results for longer lived atomic excited states involving intercombination transitions employed in the experiment with the alkaline-earth-like $^{171}\mathrm{Yb}$ atoms,  
for which $\Gamma = 48 E_{\mathrm R}$ \cite{Wang2018} is approximately 20 times smaller compared to $\Gamma = 1000 E_{\mathrm R}$ used in the present calculations. In that case,
$\operatorname{Im}{E_{\mathbf{q}}}$ would be approximately 20 smaller compared to the values presented in Fig.~\ref{fig:bulk_spectrum_N4_4panel}. 

The first four panels of Fig.~\ref{fig:bulk_spectrum_N4_4panel} compare the fine tuning point $(b,\gamma)=(1,0)$ in which (a) the effective magnetic field is non-staggered (see  Fig.~\ref{fig:magnetic-field-c-grid}(b)), with the situation shown in  (b)-(d) where the amplitude or phase deviate from the fine tuning point and thus the patches of the magnetic field of the opposite sign appear in  Fig.~\ref{fig:magnetic-field-c-grid}(b) for the effective magnetic field.
For the fine tuned configuration, $(b,\gamma)=(1,0)$, shown in Fig.~\ref{fig:bulk_spectrum_N4_4panel}(a), the energy bands  approach a set of nearly horizontal bands as $\Omega_0$ increases. 
In such a fine tuned situation, the magnetic field is non-staggered, as one can see in Fig.~\ref{fig:magnetic-field-c-grid}(b). Furthermore, since the magnetic field and the geometric scalar potential obey the AC condition, the lowest dark state band should acquire the Landau-level character discussed above and thus should be completely flat.
 
Yet, as one can see in Fig.~\ref{fig:bulk_spectrum_N4_4panel}(a), for the finite Rabi frequency $\Omega_0$ the width of the lowest band remains finite even for fine tuning when $(b,\gamma)=(1,0)$. For example, the band width equals approximately $0.13E_{\mathrm R}$, $0.051E_{\mathrm R}$ and $0,023 E_{\mathrm R}$  for relative Rabi frequencies $\Omega_0/E_{\mathrm R} = 360,\, 925\,\, \mathrm{and} \,\, 2000$, respectively, as indicated by red, blue and green arrows in Fig.~\ref{fig:bulk_spectrum_N4_4panel}(a). This is because the AC condition and the perfectly narrow lowest energy band rely on the adiabatic approximation corresponding to the infinitely large $\Omega_0$.

Panels (b)-(d) of Fig.~\ref{fig:bulk_spectrum_N4_4panel} show how the energy bands are modified when the system is moved a little away from the fine tuning $(b,\gamma)=(1,0)$, and thus the magnetic field shown in Fig.~\ref{fig:magnetic-field-c-grid}b acquires narrow patches of the opposite sign. Fig.~\ref{fig:bulk_spectrum_N4_4panel}(b) illustrates the effect of the amplitude imbalance ($b=0.9$ and $\gamma=0$), where the energy bands are broadened and acquire larger  decay rates, especially at smaller values of $\Omega_0$. On the other hand, as discussed below in more detail, around a special value of $\Omega_0$ equal to $925 E_{\mathrm R}$, the lowest Bloch band becomes much narrower compared to that for perfect tuning and the same value of $\Omega_0$.  Similar effects are produced by the phase mismatch $\gamma=0.1\pi$ and $b=1$, shown in Fig.~\ref{fig:bulk_spectrum_N4_4panel}(c), where the band broadening is larger than in Fig.~\ref{fig:bulk_spectrum_N4_4panel}(b), because the corresponding magnetic field has broader patches of opposite sign shown in blue in Fig.~\ref{fig:magnetic-field-c-grid}(b). In such a situation, band narrowing takes place at a smaller Rabi frequency $\Omega_0 = 360 E_{\mathrm R}$. The combined deviation of the amplitude and phase from the fine tuning point ($b=0.9$ and $\gamma=0.1\pi$) is shown in Fig.~\ref{fig:bulk_spectrum_N4_4panel}(d).

In the parameter regimes presented in Figs.~\ref{fig:bulk_spectrum_N4_4panel} (b)-(d) the AC matching between the scalar potential and the magnetic field still holds, but the condition $\pm B_{z}\ge 0$ is no longer fulfilled for the magnetic field. The ideal Aharonov-Bohm singularities are then replaced by finite regions of the magnetic field with the sign opposite to that of the smooth background field. These regions act as compensating flux patches that remove the exact degeneracy of the lowest Landau level in the adiabatic limit.  
Note that for no or small deviations from the fine tuning (Fig.~\ref{fig:bulk_spectrum_N4_4panel}(a)-(b)) and sufficiently large Rabi frequency $\Omega_0$, all the dark state bands are separated by energy gaps and characterized by unit Chern numbers, not only the lowest one, as in the Quantum Hall effect. 

As already noticed, a little away from the fine tuning point where $(b,\gamma)\ne(1,0)$, the lowest energy band becomes nearly degenerate at some finite value of $\Omega_0$, 
as one can see in Figs.~\ref{fig:bulk_spectrum_N4_4panel}(b)-(d). To explore this in more detail, in Figs.~\ref{fig:bulk_spectrum_N4_4panel}(e)-(f) we have plotted an enlarged spectrum  and its width  in a vicinity of the minimum bandwidth for $(b,\gamma)=(0.9,0)$ corresponding to the panel (b).

In the latter fig.~\ref{fig:bulk_spectrum_N4_4panel}(f), one can see that the minimum bandwidth  is $W_1 = 2.1\times 10^{-3}E_{\mathrm R}$ at $\Omega_0\approx 925E_{\mathrm R}$. This is about $25$ times smaller than the bandwidth  $5.1\times 10^{-2}E_{\mathrm R}$  corresponding to the perfect tuning ($b=1$) at the same Rabi frequency $\Omega_0\approx 925E_{\mathrm R}$, marked by the blue arrow in Fig.~\ref{fig:bulk_spectrum_N4_4panel}(a). 
On the other hand, the minimum bandwidth is three times smaller than the  bandwidth due to excited state decay averaged over the Brilloin zone 
$-\langle\operatorname{Im}E_{\mathbf{q}}\rangle_{\mathrm{BZ}} \approx 7\times 10^{-3}E_{\mathrm R}$ shown by the red dashed line in panel (f). These calculations have been performed for the decay rate $\Gamma =1000E_{\mathrm R}$ typical for alkali atoms. Yet, by taking $^{171}\mathrm{Yb}$ atoms with a much smaller decay rate $\Gamma = 48 E_{\mathrm R}$  \cite{Wang2018}, the radiative bandwidth reduces to $-\langle\operatorname{Im}E_{\mathbf{q}}\rangle_{\mathrm{BZ}} \approx 3.4\times 10^{-4}E_{\mathrm R}$. This is much lower than the non-radiative bandwidth $W_1$. Note that the latter $W_1$ is not sensitive to a reduction in the decay rate $\Gamma$ and remains  $W_1 = 2.1\times 10^{-3}E_{\mathrm R}$.
Further reduction of bandwidths and radiative lifetimes is possible by increasing the Rabi frequency $\Omega_0$, as we will see in the next Fig.~\ref{fig:bandwidth-N4} corresponding to $\Omega_0=2000E_{\mathrm R}$.   

\begin{figure}[tbp]
\includegraphics[width=0.95\columnwidth]{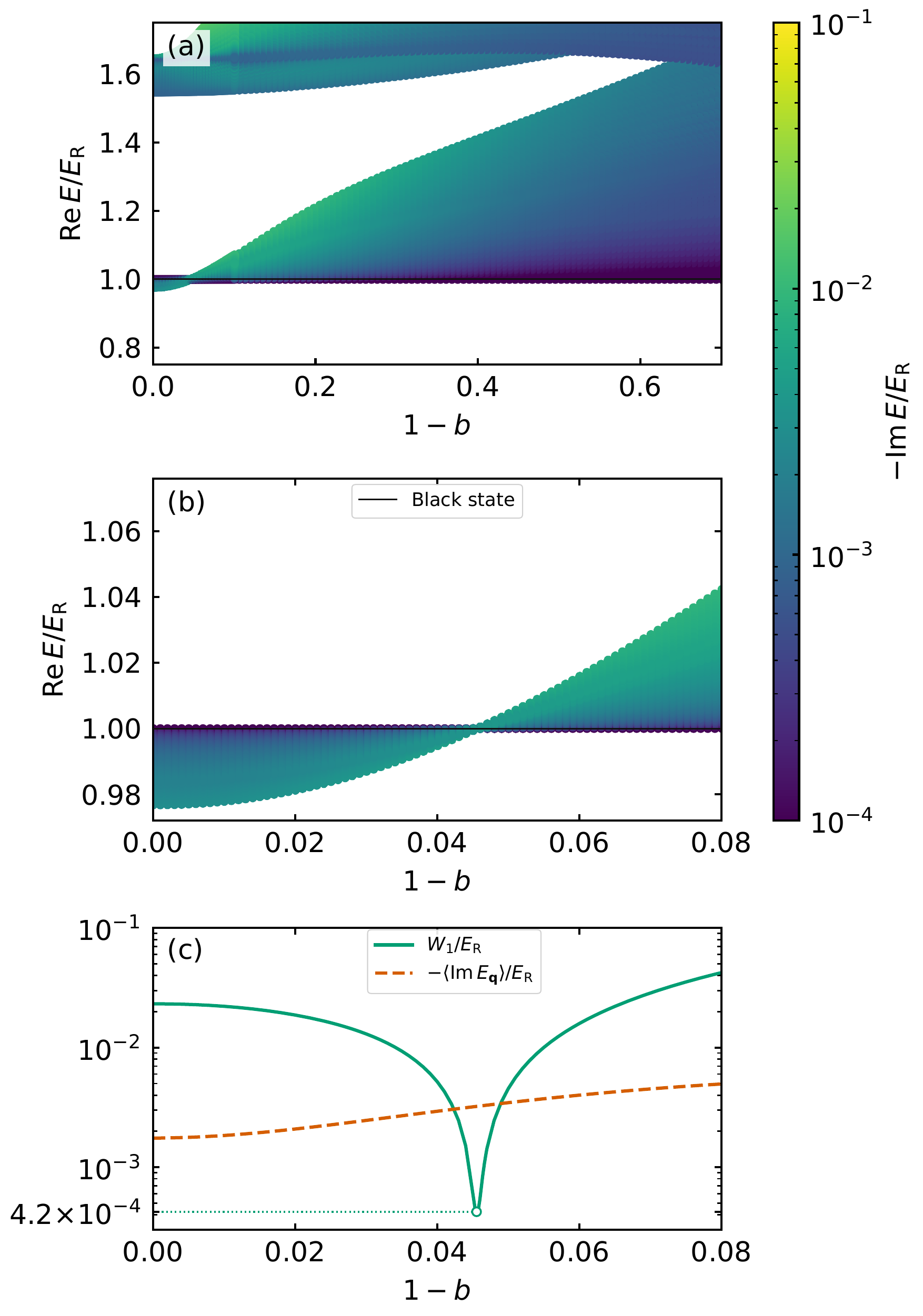}
\caption{
(a) Real part of the bulk eigen-energies versus amplitude imbalance $1-b$
for $N=4$, $\gamma=0$, $\Omega_0=\Delta=2000E_{\mathrm R}$, and
$\Gamma =1000E_{\mathrm R}$.
(b) Enlarged lowest-band spectrum for $0.92\leq b\leq1$.
(c) The band width $W_1/E_{\mathrm R}$ (solid green) and the band averaged imaginary part of the energy
$-\langle\operatorname{Im}E_{\mathbf{q}}\rangle_{\mathrm{BZ}}/E_{\mathrm R}$ (dashed orange).
The minimum band width of  $4.23505\times10^{-4}E_{\mathrm R}$
at $b\simeq0.95446347$, is marked on the logarithmic axis in (c).
Colors in (a)-(b) show $-\operatorname{Im}E/E_{\mathrm R}$;  the lossless black state at $E/E_{\mathrm R}=1$ is
drawn in black above the colored spectrum.
The calculations have been carried out using $N_{\mathrm{F}}=21$ Fourier modes in both spatial directions  $x$ and $y$. The convergence of the spectra has been checked. 
}
\label{fig:bandwidth-N4}
\end{figure}

To look at things from a somewhat different angle, in Fig.~\ref{fig:bandwidth-N4}(a) we have plotted the dependence of the bulk spectrum on the amplitude imbalance $1-b$ for $\gamma =0$, a fixed value of the Rabi frequency  $\Omega_0=2000E_{\mathrm R}$ and a moderate positive detuning ($\Delta_0=\Omega$) to reduce losses, as in previous theoretical and experimental treatments on dark state atoms \cite{Zoller2016,Wang2018,Nascimbene25PRL,Burba25PRR}.  

Fig.~\ref{fig:bandwidth-N4}(b) shows the enlarged spectrum in a vicinity of the minimum bandwidth for $0.92\le b \le 1$ and panel (c) presents the corresponding relative bandwidth $W_1\,/E_{\mathrm R}$.
 One can see that reducing $b$ from $1$ (perfect tuning, completely non-staggered magnetic field) to $b \approx 0.954$ (weakly staggered magnetic field), the width of the lowest energy band $W_1$ reduces around 50 times from $W_1 = 2.3\times 10^{-2}E_{\mathrm R}$  at $b=1$ (green arrows in Fig.~\ref{fig:bulk_spectrum_N4_4panel}(a)) to $W_1 = 4.24\times 10^{-4}E_{\mathrm R}$  at $b\simeq0.95446$ in Fig.~\ref{fig:bandwidth-N4}(c). On the other hand, the radiative width $-\langle\operatorname{Im}E_{\mathbf{q}}\rangle_{\mathrm{BZ}}$ shown by the orange dashed line in Fig.~\ref{fig:bandwidth-N4}(c) is now around 7.6 times larger than the non-radiative bandwidth $W_1$. Yet, by decreasing the excited state decay rate from $\Gamma =1000E_{\mathrm R}$ to $\Gamma = 48 E_{\mathrm R}$ (corresponding to the $^{171}\mathrm{Yb}$ atoms used in the experiment  \cite{Wang2018}), the radiative linewidth reduces below the non-radiative width $W_1$ to $-\langle\operatorname{Im}E_{\mathbf{q}}\rangle_{\mathrm{BZ}}\approx 1.6\times 10^{-4}E_{\mathrm R}$, and the corresponding averaged lifetime of the band states is around 0.15 s.
 As one can see from the dashed orange line in  Fig.~\ref{fig:bandwidth-N4}(c), the radiative widths can be made a few times smaller by going from $b=0.955$ to the fine tuning where $b=1$, so the radiative lifetimes of the band states can be increased to around $1\mathrm{s}$. The negative side of this is that the non-radiative width of the lowest dark state band $W_1 = 2.3\times 10^{-2}E_{\mathrm R}$  would then be around 50 times larger than that for $b=0.955$. The bands can be further narrowed and the radiative lifetimes can be further increased by taking larger values of the Rabi frequency $\Omega_0$. The Rabi frequency $\Omega_0=2000E_{\mathrm R}$ used here is about 1.5 times smaller than the one employed in the experiment \cite{Wang2018}, so some increase in the Rabi frequency is possible.

The experiment~\cite{Wang2018} dealt with one-dimensional (1D) dark state lattices with subwavelength barriers, and  the observed lifetime of the lowest band states was $44\,\mathrm{ms}$. The current theoretical analysis indicates that in two dimensional dark state lattices this number can be larger (of the order of $1\,\mathrm{s}$) for a similar laser power. An increase of the lifetime of the lowest-band atoms in the two dimensional setting can be explained using the following arguments.
For both 1D \cite{Zoller2016,Jendrzejewski2016,Wang2018} and 2D \cite{Gvozdiovas23PRA,Burba25PRR} dark-state lattices, the geometric scalar potential $\phi$ provides the subwavelength potential barriers of width $x_{0}$ and height proportional to $1/x_{0}^{2}$. In 1D systems, the area over such a barrier 
is proportional to $1/x_{0}$ and thus diverges for $x_{0}\rightarrow 0$, while in 2D systems the integral has a finite value in the limit where $x_{0}\rightarrow 0$. The geometric scalar potential $\phi\propto\left|\mathbf{A}_{{\rm DB}}\right|^{2}$ is determined by the off diagonal matrix element of the vector potential $\mathbf{A}_{{\rm DB}}$, and the latter $\mathbf{A}_{{\rm DB}}$ describes also non-adiabatic transitions leading to occupation of the lossy excited state. Consequently, the 1D subwavelength lattices experience considerable losses at the narrow (subwavelength) barriers produced by the geometric scalar potential.

The narrower the barriers, the larger the losses. The losses limit the width of the 1D subwavelength barriers, which is proportional to the dimensionless parameter $\epsilon$ \cite{Wang2018} corresponding to $1-b$ in our 2D system. In ref.~\cite{Wang2018}, $\epsilon=1/7$ was used; lower values of $\epsilon$ would lead to a considerable increase in losses. In contrast, in the 2D setting, the integral over the geometric potential does not diverge, so the losses decrease with $1-b$ and reach the minimum at the fine tuning point where $b=1$, as can be seen from the orange dashed line in Fig.~\ref{fig:bandwidth-N4} (c) representing the losses for $\Gamma = 1000 E_{\mathrm{R}}$. Yet, this is at the cost of the considerably larger non-radiative bandwidth
at the fine tuning point $b=1$ compared to the minimum non-radiative bandwith further away from this point, as discussed above.

\subsection{Origin of band narrowing}
\label{Eluciation}

The significant reduction of the width of the lowest dark state band at the special values of $\Omega_0$ and $(b,\gamma)$ observed in Figs.~\ref{fig:bulk_spectrum_N4_4panel}-\ref{fig:bandwidth-N4}, is associated with a delicate balance between two imperfections, namely non-adiabatic corrections due to the finite Rabi frequency $\Omega_0$ and deviation from the fine tuned point $(b,\gamma)=(1,0)$, at which the 
 lowest dark state Bloch band is flat in the adiabatic limit.

 Before presenting a more detailed mechanism underlying the band narrowing, it is helpful to recall that the system contains the black state [given by Eq.~\eqref{eq:Black}] which is immune to non-adiabatic coupling and is an exact eigenstate of the full Hamiltonian $\hat{H}$ with eigenenergy equal to the recoil energy $E_{\mathrm R}\equiv E_{\kappa}$
 for any values of $\Omega_0$, $b$ and $\gamma$ defining the Rabi frequencies $\Omega_{1,2}$  in Eqs.~(\ref{eq:Omega_1-four-plane waves})-(\ref{eq:Omega_2-four-plane waves}). Importantly, the black state  is a Bloch state with zero quasimomentum $\mathbf{q}$ in an extended Brillouin zone \footnote{In an extended Brillouin zone, the Bloch states are eigenstates of a combined translation operator involving a spatial translation by a half of the lattice period along both Cartesian axes,  $\left(\mathbf{e}_x\pm\mathbf{e}_y\right)\pi/\kappa$, and a unitary operation $\left|e\right\rangle \left\langle e\right| -\left|1\right\rangle \left\langle 1\right| -\left|2\right\rangle \left\langle 2\right|$, as in refs.~\cite{Gvozdiovas23PRA,Burba25PRR}.}
 \begin{equation}
 \left|\mathrm{Black}\right\rangle = \left|\psi_{\mathbf{q}=0}\right\rangle\,,\quad E_{\mathbf{q}=0}=E_{\kappa}\,.
 \label{Black-phi-q zero}
 \end{equation}
Consider first the adiabatic picture relevant to sufficiently large values of the Rabi frequency $\Omega_0$, for which the atomic dynamics is confined within the dark state manifold.
  In that case, for $(b,\gamma)=(1,0)$ 
the lowest Bloch band of the dark state manifold is completely degenerate and characterized by the eigen-energy $E_{\kappa}$ because of the AC condition and the non-staggered magnetic field. 
For $(b,\gamma)\ne(1,0)$, the patches of the magnetic field of opposite sign and the corresponding scalar potential emerge (Fig.~\ref{fig:magnetic-field-c-grid}(b)),  shifting upward the energies of all Bloch states except that with $\mathbf{q}=0$ (the black state), which retains the original energy $E_{\kappa}$ and thus becomes the lowest band state. 
This is clearly seen in Figs.~\ref{fig:bulk_spectrum_N4_4panel}-\ref{fig:bandwidth-N4}, in which the band minimum is strictly at the energy of the lossless black state $E_{\kappa}$ for sufficiently large Rabi frequency $\Omega_0$ (Fig.~\ref{fig:bulk_spectrum_N4_4panel}) or for sufficiently large deviation from the fine tuning (Fig.~\ref{fig:bandwidth-N4}), i.e. when the deviation from the fine tuning has a bigger effect on the spectrum than the non-adiabatic corrections.

Now, let us examine the additional effects of non-adiabatic corrections  that are the most important in the spatial areas of small Rabi frequencies where $\Omega\ll\Delta$.  In such a situation, the projected description of the atom-light coupling in terms of the dark and bright states by  Eqs.~\eqref{eq:V-eff}-\eqref{eq:V_0} is relevant.
For the positive detuning ($\Delta>0$) considered here, the light shift given by  Eqs.~\eqref{eq:V-eff}-\eqref{eq:V_0}, is positive, which places the bright-state energy above the dark-state energy. Thus, the non-adiabatic coupling between the dark- and bright-state manifolds pushes downward the states of the lowest dark-state band except that with $\mathbf{q}=0$ (the black state), 
as the latter is immune to the non-adiabatic coupling. This leads to lowering of the energy of the Bloch states with $\mathbf{q}\ne0$ and thus the band flattening until the bandwidth reaches the minimum for a certain value of Rabi frequency $\Omega_0$ and fixed $b$ (Fig.~\ref{fig:bulk_spectrum_N4_4panel}) or a certain value $b$ and fixed $\Omega_0$ Fig.~(\ref{fig:bandwidth-N4}).  A finite (yet small) minimum bandwidth appears because the energies of band states with $\mathbf{q}\ne0$ reach the energy  $E_{\kappa}$ of the black ($\mathbf{q}=0$) state for slightly different combinations of $\Omega_0$ and $b$.

 In this way, the minimum bandwidth relies on the balance between the effects due to the deviation from the perfect tuning and non-adiabatic corrections.  To reach the minimum bandwidth, larger deviations from the perfect tuning $(b,\gamma)=(1,0)$ require larger non-adiabatic corrections and hence smaller values of $\Omega_0$ or larger detuning $\Delta$.  This can be seen in Figs.~\ref{fig:bulk_spectrum_N4_4panel}-\ref{fig:bandwidth-N4} with different Rabi frequencies $\Omega_0$ and the same detuning $\Delta=2000E_{\mathrm R}$.   On the other hand, keeping the same $\Omega_0=2000E_{\mathrm R}$ and increasing the detuning, $\Delta/E_{\mathrm R}=1400$, $2000$ and $3000$, the minimum bandwidth occurs at an increasing deviation from the fine tuning point: $1-b\simeq0.038$, $0.045$ and $0.055$ for  $\gamma=0$.
 At these points, the minimum non-radiative bandwidths increase and equal to $W_{1,\min}/E_{\mathrm R}\simeq3 \times10^{-4}$, $4.2 \times10^{-4}$ and $6.29\times10^{-4}$, respectively. Yet, the corresponding averaged radiative line-widths decrease with the detuning and are equal, respectively, to $-\langle\operatorname{Im}E_{\mathbf q}\rangle_{\mathrm{BZ}}/E_{\mathrm R} \simeq 1.9\times10^{-4}$, $1,6\times10^{-3}$ and $1.3\times10^{-4}$  for $\Gamma=48E_{\mathrm R}$.  This is because, as discussed at the end of Sec.~\ref{Bulk spectrum}, larger values of $1-b$ mean larger spots of subwavelength barriers of the geometric scalar potential, causing larger radiative losses.

After passing the point of minimum bandwidth, the $\mathbf{q}=0$ state becomes the highest energy state of the band. This is seen in Figs.~\ref{fig:bulk_spectrum_N4_4panel}- \ref{fig:bandwidth-N4}, in which the straight horizontal line corresponding to the black ($\mathbf{q}=0$) state, changes from the minimum to the maximum of the band as $\Omega_0$ (Fig.~\ref{fig:bulk_spectrum_N4_4panel}) or $1-b$ (Fig.~\ref{fig:bandwidth-N4}) passes through the region of minimum bandwidth. In this regime the non-adiabatic effects prevail over the deviation from the fine tuning. For sufficiently small Rabi frequencies ($\Omega_0 = \Delta/10\lesssim 100 E_{\mathrm R}$ in Fig.~\ref{fig:bulk_spectrum_N4_4panel}), the non-adiabatic transitions  between the dark and bright states are so strong that the individual Bloch bands are merged. 
A detailed analysis of the combined influence of non-adiabatic effects and the deviation from the fine tuning point on the dark state spectrum will be presented in a separate study. 

In the next Subsection~\ref{Existence of ideal} we will consider the geometry of the lowest dark state band. We will demonstrate that the nearly degenerate lowest energy band forming for finite $\Omega_0$ and some departure from the fine tuned point $(b,\gamma)=(1,0)$  observed in Figs.~\ref{fig:bulk_spectrum_N4_4panel}-\ref{fig:bandwidth-N4}, is characterized by nearly perfect band geometry. This can be used for simulating the fractional Quantum Hall effect if the atom-atom interaction is added.   

\subsection{Band geometry}
\label{Existence of ideal}

Recently, there has been considerable interest in simulating fractional
Chern insulators (FCIs) using  ultracold
atoms \cite{Cooper-Dalibard13PRL,Greiner23Nature,Joachim24PRL}. Although
having a non-zero Chern number is a necessary condition for the formation
of FCI, it is not sufficient to stabilize the many-body FCI phase
\cite{Wang21PRL,Crepel2023PRR,Ledwith23PRB}. Hence, the energy bands
must satisfy certain conditions which allow a mapping to Landau levels. This can be studied using the quantum geometric tensor (QGT) $\mathcal{Q}_{\mathbf{q}}^{ab}$,
given by \cite{Mitscherling2025PRB}: 

{\renewcommand{\theequation}{\textcolor{red}{\arabic{equation}}}
\begin{equation}
\mathcal{Q}_{\mathbf{q}}^{ab}=\sum_{s\in\mathcal{S}}\langle\partial^{a}u_{\mathbf{q}s}|\hat{R}_{\mathbf{q}}|\partial^{b}u_{\mathbf{q}s}\rangle\,,\label{eq:Q_q^ab}
\end{equation}
}
where $\partial^{a}$ and $\partial^{b}$, with $a,b\in\left\{ q_{x},q_{y}\right\} $, label the partial derivatives 
with respect to the quasi-momentum components.
Here $|u_{\mathbf{q}s}\rangle$
is the periodic Bloch state vector at the quasi-momentum $\mathbf{q}$ for the $s$-th band, $\mathcal{S}$ is the set of bands under consideration,
the operator $\hat{R}_{\mathbf{q}}=\hat{I}-\sum_{s\in\mathcal{S}}|u_{\mathbf{q}s}\rangle\langle u_{\mathbf{q}s}|$
projects onto the complement subspace, and $\hat{I}$ is the identity
operator. No summation over $s$ appears if one considers
a geometric tensor associated only with a single Bloch band $s$.

Note that the form of QGT presented in Eq.~(\ref{eq:Q_q^ab}) implies a continuous phase gauge
for the Bloch state vectors, which is not convenient
for numerical calculations. Therefore, we will use the QGT expressed in terms
of gauge-invariant projectors as \cite{Mitscherling2025PRB}:

{\renewcommand{\theequation}{\textcolor{red}{\arabic{equation}}}
\begin{equation}
\mathcal{Q}_{\mathbf{q}}^{ab}=\mathrm{Tr}\left(\hat{P}_{\mathbf{q}}\left(\partial^{a}\hat{P}_{\mathbf{q}}\right)\left(\partial^{b}\hat{P}_{\mathbf{q}}\right)\right)\,,\label{eq:Q_q^ab-alternative}
\end{equation}
}
where $\hat{P}_{\mathbf{q}}= \hat{I} - \hat{R}_{\mathbf{q}}=\sum_{s\in\mathcal{S}}|u_{\mathbf{q}s}\rangle\langle u_{\mathbf{q}s}|$
is the projector onto the subspace of the relevant Bloch bands.

We will use two well-established quasi-momentum space criteria
for $\mathbf{q}$-ideal Chern bands \cite{Crepel2023PRR}:

(1) The determinant of the integral of the QGT over the Brillouin zone is equal to zero:
\begin{equation}
\begin{gathered}
 D_{{\rm QGT}}\coloneqq\det\langle\mathcal{Q}_{\mathbf q}\rangle_{\mathrm{BZ}}=0\,,\\
 \langle\mathcal{Q}_{\mathbf q}\rangle_{\mathrm{BZ}}
 \coloneqq\frac{1}{A_{\mathrm{BZ}}}\int_{\mathrm{BZ}}d^2\mathbf q\,\mathcal{Q}_{\mathbf q}\,.
\end{gathered}
 \label{eq:D_QGT}
\end{equation}
where $A_{\mathrm{BZ}}$ is the Brillouin-zone area and $\mathcal{Q}_{\mathbf q}$ is a $2\times2$ matrix with elements $\mathcal{Q}_{\mathbf q}^{ab}$. The normalized average matches the numerical convention; its determinant vanishes under the same condition as the determinant of the unnormalized integral.

(2) The QGT has a constant ($\mathbf{q}$-independent)
null vector {$w_\mathbf{q} $} for all quasi-momenta $\mathbf{q}$:
\begin{equation}
\mathcal{Q}_{\mathbf{q}}w_{\mathbf{q}}=0\,,\quad\mathrm{with}\quad w_\mathbf{q}\equiv w\,,
 \label{eq:Null vector}
\end{equation}
where the null vector $w_{\mathbf{q}}$ is a $2\times 1$ (column) matrix normalized to unity: $w_{\mathbf{q}}^{\dagger}w_{\mathbf{q}}=1$.

\begin{figure}[t]
\centering
\includegraphics[width=0.5\textwidth]{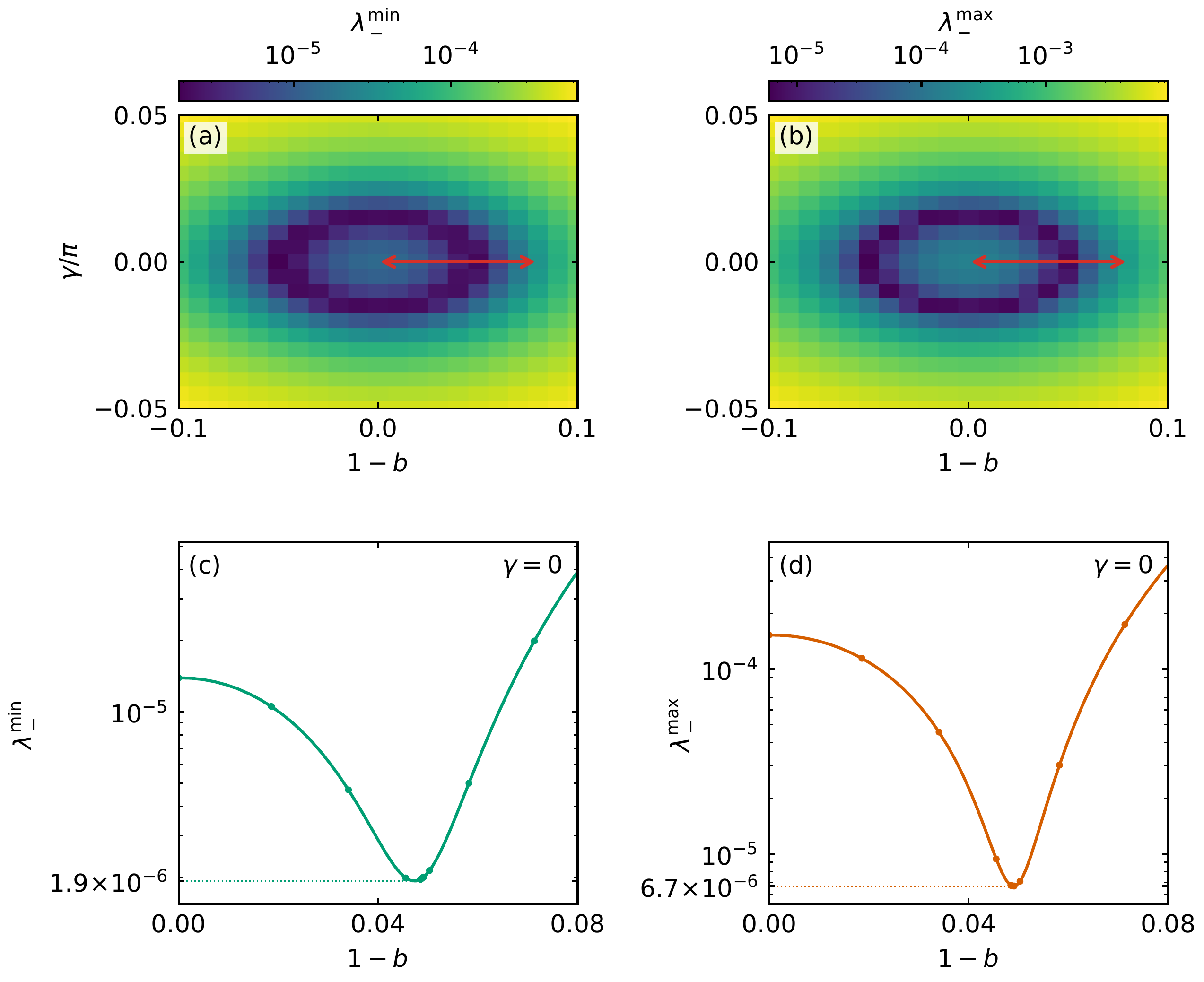}
\caption{
(a,b) The minimum and maximum of the lowest eigenvalue of QGT $\lambda_-(\mathbf q)$  over the sampled
Brillouin zone, $\lambda_-^{\mathrm{min}}$ and $\lambda_-^{\mathrm{max}}$, plotted against $1-b$ and $\gamma/\pi$.
(c,d) The corresponding $\gamma=0$ sections, along the red arrows in
(a,b). The calculations have been carried out using $N_{\mathrm{F}}=21$ Fourier modes in both spatial directions $x$ and $y$ and, the convergence of the results has been checked. 
}
\label{fig:QGT-eigenvalue-ranges}
\end{figure}

 For a generic non-ideal band, an exact null vector  $w_{\mathbf{q}}$ does not need to exist. 
 To deal with the second requirement \eqref{eq:Null vector},  we therefore select $w_{\mathbf q}$ as the normalized eigenvector with the
smaller eigenvalue: $\mathcal Q_{\mathbf q}w_{\mathbf q}
=\lambda_-(\mathbf q)w_{\mathbf q}$, and check how close  to zero the eigenvalue $\lambda_-(\mathbf q)$ is.
Figure~\ref{fig:QGT-eigenvalue-ranges} shows the minimum and maximum of
$\lambda_-(\mathbf q)$ in the sampled Brillouin zone, together with the corresponding  cuts in $\gamma=0$. At fine tuning $(b,\gamma)=(1,0)$ the eigenvalue $\lambda_-(\mathbf q)$
ranges from $\lambda_-^{\mathrm{min}}=1.39\times10^{-5}$ to $\lambda_-^{\mathrm{min}}= 1.53\times10^{-4}$, with mean value
$\langle\lambda_-(\mathbf q)\rangle = 5.34\times10^{-5}$. At $(b,\gamma)=(0.95,0)$ the range is
$[2.10\times10^{-6},6.93\times10^{-6}]$, with mean value $3.54\times10^{-6}$.  Hence, there is an order of magnitude reduction in the minimum eigenvalues $\lambda_-(\mathbf q)$ going from fine tuning $(b,\gamma)=(1,0)$ to $(b,\gamma)=(0.95,0)$, and the latter point corresponds nearly exactly to the minimum bandwidth in Fig.~\ref{fig:bandwidth-N4}(c).
 Similarly, the maximum ratio between the lower and upper eigenvalues $\lambda_-/\lambda_+$ decreases from
$\lambda_-/\lambda_+\simeq 10^{-3}$ to $\lambda_-/\lambda_+\simeq 4.5\times10^{-5}$ when going to the fine tuning point to $(b,\gamma)=(0.95,0)$. 
In this way, the minimum eigen-value vector $w_{\mathbf{q}}$ can be considered a null vector with good accuracy, especially a little away from the fine tuning point $(b,\gamma)=(1,0)$ around the minimum bandwidth.

To quantify the second requirement \eqref{eq:Null vector}, we introduce a projector onto the vector with minimum eigen-value, $\Pi_{\mathbf q}$, its average $\overline\Pi$ and the  variance  over the Brillouin zone $\sigma_P^2$: 
\begin{equation}
 \Pi_{\mathbf q}=w_{\mathbf q}w_{\mathbf q}^{\dagger}
 \,\,,
 \qquad \overline\Pi=\langle\Pi_{\mathbf q}\rangle_{\mathrm{BZ}},
 \label{eq:QGT-direction-projector}
\end{equation}
\begin{equation}
 \sigma_P^2
 =\left\langle\|\Pi_{\mathbf q}-\overline\Pi\|_F^2
   \right\rangle_{\mathrm{BZ}} 
 =1-\operatorname{Tr}\overline\Pi^2\,,
 \label{eq:sigma_QGT}
\end{equation}
where $\|\ldots\|$ signifies the Frobenius norm of a matrix: $\|M\|_F^2=\operatorname{Tr}(M^\dagger M)$. The  variance $\sigma_P^2$ 
is invariant under the transformation 
$w_{\mathbf q}\mapsto e^{i\theta(\mathbf q)}w_{\mathbf q}$, so one does not need to worry about the gauge choice for the null vector $w_{\mathbf q}$.
Small values of the spread ($\sigma_P=\sqrt{\sigma_P^2}\ll1$) imply a nearly constant minimum-eigenvalue vector $w_{\mathbf q}$ over the Brillouin zone, up to its arbitrary phase.

\begin{figure}[tbp]
\includegraphics[width=0.95\columnwidth]{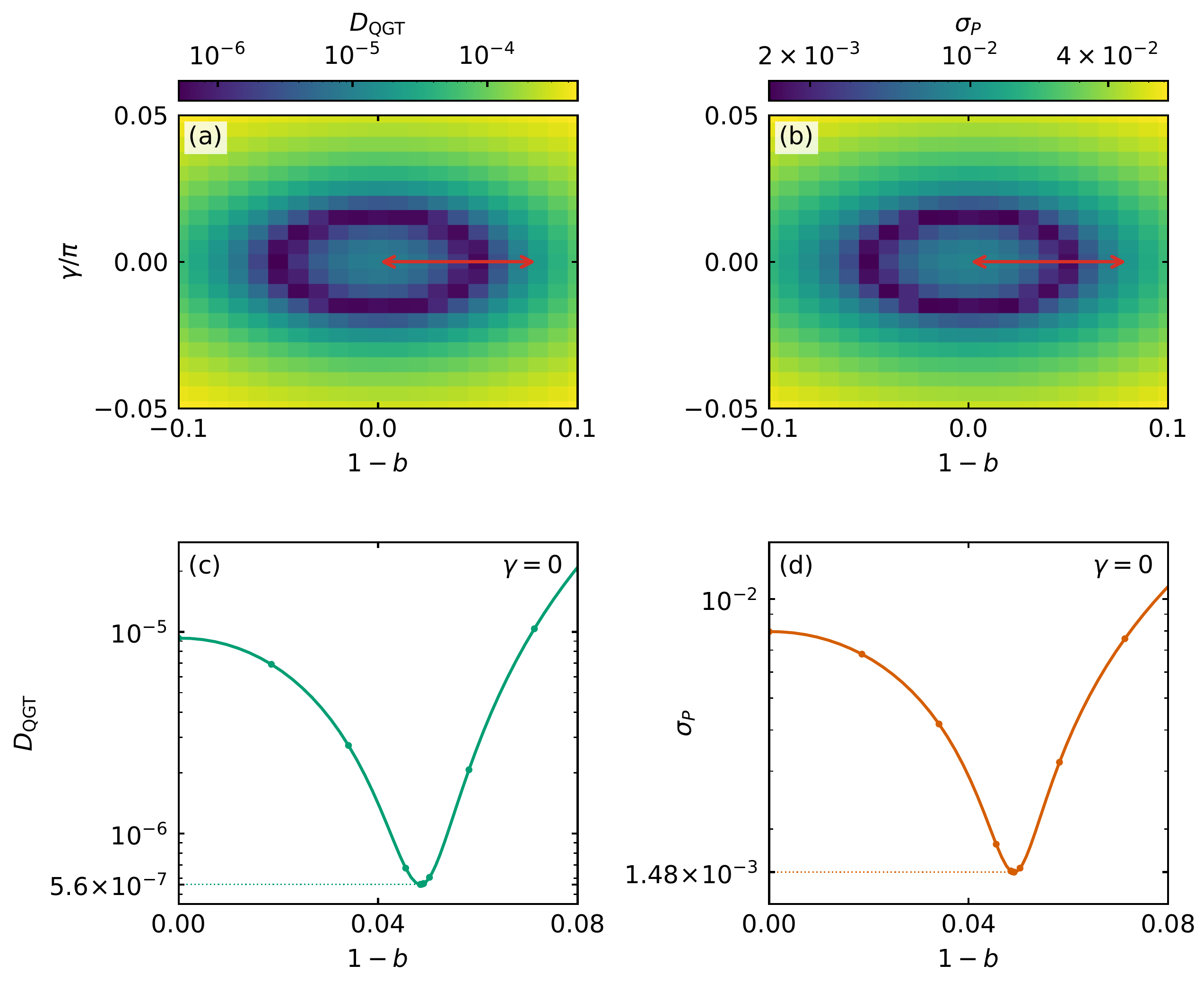}
\caption{Quantum geometry of the lowest energy band
for $N=4$ and $\Omega_0 = \Delta =2000E_{\mathrm R}$ in a vicinity of the fine tuning point $(b,\gamma)=(1,0)$.  (a) The determinant of the Brillouin-zone-averaged QGT given by Eq.~\eqref{eq:D_QGT} and (b) the spread of the minimum-eigenvalue direction $\sigma_P$ defined by Eq.~\eqref{eq:sigma_QGT} in the $b$-$\gamma$ plane. (c,d) The corresponding refined cuts at $\gamma=0$ over $0\leq1-b\leq0.08$,
indicated by red double-headed arrows in the upper panels (a)-(b).
All horizontal axes show $1-b$.  The calculations have been carried out using $N_{\mathrm{F}}=21$ Fourier modes in both spatial directions  $x$ and $y$, and the convergence of the results has been checked.}

\label{fig:QGT-phase-diag}
\end{figure}

Figure~\ref{fig:QGT-phase-diag} shows the quantities $D_{{\rm QGT}}$ and  $\sigma_P$ that characterize the quantum geometry of the lowest energy band of the dark state manifold
for $N=4$ and $\Omega_0=2000E_{\mathrm R}$ around the fine-tuned point where $(b,\gamma)\approx(1,0)$. At the fine-tuning point corresponding to $(b,\gamma) = (1,0)$, we find  $D_{{\rm QGT}}\approx 0.9 \times 10^{-5}$ and $\sigma_{{\rm QGT}}\approx 0.8 \times 10^{-2}$.
This is much less than unity, but still is non-zero, because for finite $\Omega_0$ a deviation from the strictly adiabatic description remains.
In Figs.~\ref{fig:QGT-phase-diag} (a)-(b) one can see that the fine tuned point $(b,\gamma)=(1,0)$ is surrounded by an ellipse (similar to the one in Fig.~\ref{fig:QGT-eigenvalue-ranges}) on which  $D_{{\rm QGT}}$ and $\sigma_{{\rm QGT}}$ are characterized by considerably lower values than at the fine tuned point. The elipse corresponds approximately to the values of $(b,\gamma)$ where the width of the lowest band is minimum.
In particular, along the refined cuts $\gamma=0$ presented in Figs.~\ref{fig:QGT-phase-diag} (c)-(d), the minima  $D_{{\rm QGT}}\approx 5.6\times10^{-7}$ and  $\sigma_{{\rm QGT}}\approx 1.48\times10^{-3}$ appear at $b\approx 0.95$, almost at the same place as the minimum bandwidth in Fig.~\ref{fig:bandwidth-N4}. These values of $D_{{\rm QGT}}$ and $\sigma_{{\rm QGT}}$ are an order of magnitude smaller than those at the fine tuning point where $b=1$. 
 In this way, for the finite atom-light coupling strength $\Omega_0$, a proper deviation from the fine tuned point $(b,\gamma)=(1,0)$ allows one to considerably reduce the width of the lowest dark-state band and additionally have closer to ideal band geometry, which is needed to simulate fractional Quantum Hall states.

\section{Concluding remarks}\label{sec:Discussion}

We investigated the realization of flat and topologically nontrivial energy bands for ultracold atoms adiabatically following the dark state in the $\Lambda$-type atom--light coupling scheme under the Aharonov--Casher (AC) condition. The AC condition provides a  relation between the geometric scalar potential and the synthetic magnetic field, allowing the emergence of a fully degenerate lowest-Landau-level-like band in the presence of spatially inhomogeneous magnetic fields, as long as it has the proper sign.
A general criterion for the AC condition for the $\Lambda$ atom--light coupling was derived, showing that the AC condition can be achieved when the Rabi frequencies are constructed from properly tuned superpositions of plane waves. In particular, a fine-tuned setup composed of three, four, or six ($N=3,\, 4,\,6$) plane waves was shown to generate a smooth synthetic magnetic field together with arrays of non-measurable Aharonov--Bohm flux singularities, leading to the completely flat lowest energy band in the adiabatic limit corresponding to an infinitely large strength of the atom-light coupling.
We further analyzed the effects of deviations from fine tuning of laser fields and finite atom--light coupling strength. Imperfect tuning broadens the singular magnetic fluxes into narrow regions of magnetic field of the sign opposite to the background magnetic field, leading to dispersion in the lowest energy band. Similarly, non-adiabatic corrections associated with finite coupling strength also contribute to band broadening. Importantly, we showed that these two imperfections can compensate for each other, resulting in an almost completely flat lowest band with nearly ideal band geometry needed to simulate the fractional Quantum Hall states.
These findings may facilitate the experimental realization and exploration of strongly correlated topological phases, including fractional Quantum Hall states, in highly controllable ultracold atomic systems.

\section*{Acknowledgments}
This work was supported by the Research Council of Lithuania
(Grant No. S-MIP-24-97).
~GJ thanks Nigel Cooper for helpful discussions on the AC condition for the Lambda scheme involving three symmetrically distributed plane waves \cite{Sommer-Cooper2509.01481} and 
for drawing attention to the comment presented in ref.~\cite{Sommer-Cooper2509.01481} that the AC condition can also be fulfilled for four symmetrically distributed fine tuned plane waves employed in the revised version of ref.~\cite{Nascimbene25PRL}. D.B. used resources at the High Performance Computing Center (HPCC), ``HPC Saul\.etekis'' in Vilnius University, Faculty of Physics. D.B. also utilized the computing resources of the Scientific Computing and Data Analysis section of Core Facilities at Okinawa Institute of Science and Technology Graduate University (OIST).

\newpage

\bibliography{2D_Dark_state_lattices}

\end{document}